\documentclass[lettersize,journal]{IEEEtran}
\usepackage{amsmath,amsfonts}
\usepackage[linesnumbered, ruled]{algorithm2e}
\SetKwRepeat{Do}{do}{while}
\usepackage{array}
\usepackage{textcomp}
\usepackage{stfloats}
\usepackage{url}
\usepackage{verbatim}
\usepackage{graphicx}
\usepackage{color}
\usepackage{multirow}
\usepackage{subcaption}
\usepackage{booktabs}
\usepackage{bbding}
\usepackage{threeparttable}
\usepackage{cite}
\usepackage[pagebackref,breaklinks,colorlinks,bookmarks=false]{hyperref}

\begin{document}
	
	\title{Exploring Long- and Short-Range Temporal Information for Learned Video Compression}
	
	\author{Huairui Wang, \emph{Student Member, IEEE} and Zhenzhong 		Chen\textsuperscript{*\thanks{This work was supported in part by National Natural Science Foundation of China (Grant No. 62036005). The numerical calculations in this paper have been done on the supercomputing system in the Supercomputing Center of Wuhan University. (* Corresponding author: Zhenzhong Chen.)}\thanks{Huairui Wang and Zhenzhong Chen are with the School of Remote Sensing and Information Engineering, Wuhan University, Wuhan 430079, China (e-mail: wanghr827@whu.edu.cn; zzchen@ieee.org).}}, \emph{Senior Member, IEEE}
	}
	
	
	\maketitle
	\vspace{-25pt}
	\begin{abstract}
		Learned video compression methods have gained various interests in the video coding community. Most existing algorithms focus on exploring short-range temporal information and developing strong motion compensation. Still, the ignorance of long-range temporal information utilization constrains the potential of compression. In this paper, we are dedicated to exploiting both long- and short-range temporal information to enhance video compression performance. Specifically, for long-range temporal information exploration, we propose a temporal prior that can be continuously supplemented and updated during compression within the group of pictures (GOP). With the updating scheme, the temporal prior can provide richer mutual information between the overall prior and the current frame for the entropy model, thus facilitating Gaussian parameter prediction. As for the short-range temporal information, we propose a progressive guided motion compensation to achieve robust and accurate compensation. In particular, we design a hierarchical structure to build multi-scale compensation, and by employing optical flow guidance, we generate pixel offsets as motion information at each scale. Additionally, the compensation results at each scale will guide the next scale’s compensation, forming a flow-to-kernel and scale-by-scale stable guiding strategy. Extensive experimental results demonstrate that our method can obtain advanced rate-distortion performance compared to the state-of-the-art learned video compression approaches and the latest standard reference software in terms of PSNR and MS-SSIM. The codes are publicly available on: https://github.com/Huairui/LSTVC.
	\end{abstract}
	
	
	\section{Introduction}
	Nowadays, the increasing demand for higher resolution, frame rate, and dynamic range video content makes the need for efficient video compression methods becomes more urgent. The widely adopted traditional video codecs, such as AVC \cite{h264}, HEVC \cite{HEVC} and VVC \cite{bross2021developments}, have achieved promising rate-distortion performance as they have been continuously improved by experts worldwide in recent decades. Still, they are encountering limitations in terms of optimization schemes and optimization objectives. Traditional video codecs are designed in a hybrid style that heavily relies on hand-crafted techniques to improve compression efficiency. Their sub-modules are well designed individually but unable to be optimized simultaneously, thus limiting the compression potential. Besides, it is hard to optimize the hybrid coding framework towards complex objective metrics such as MS-SSIM \cite{wang2003multiscale} and LPIPS \cite{zhang2018unreasonable}. In summary, strong requirements exist to explore new video coding frameworks and paradigms as potential candidates for future video codec, considering the unique characteristics of video content and the rapid development of deep learning technologies.
	
	Recent learning-based methods for video compression have attracted significant interest in video processing and coding communities. Some attempts have been made in the pioneer work \cite{Lu2019CVPR,yang2020learning} to optimize the video compression framework in an end-to-end fashion. For the entropy model optimization in recent video compression methods \cite{Lu2019CVPR,lu2020end,rippel2019learned,liu2020learned,hu2021fvc}, most of them directly adopt hyperprior \cite{balle2018variational} or auto-regressive prior \cite{minnen2018joint} which are proposed for image compression. These entropy models are dedicated to reducing statistical dependencies and fit the actual marginal distribution of each input image with the predicted distribution. However, there are few methods to design special prior information for entropy models based on the content characteristics of videos. According to our experiments, taking advantages of the inter-frame spatial correlation, the temporal prior can allow for significant performance gains while balancing model efficiency.
	
	As for motion compensation, early solutions \cite{Lu2019CVPR,rippel2019learned} abandoned the block-based motion estimation in the traditional method and chose optical flow to capture the temporal information. The reconstruction quality and bits consumption depend on the optical flow prediction accuracy. In other words, the flow-based compression methods rely on a pixel-to-pixel mapping scheme to model the motion information. In that case, inaccurate flow estimation may result in the weird reconstruction artifact during the pixel domain warping \cite{tian2020tdan}. Deformable convolution has been applied in FVC \cite{hu2021fvc} for feature alignment, achieving more effective motion compensation. Nevertheless, without appropriate constraints for the training of deformable compensation, it may lead to an overflow in offset prediction and cause sub-optimal compensation performance.
	
	To alleviate the problems caused by the absence of video-oriented prior and unstable motion compensation, we propose a novel framework for exploring Long- and Short-range Temporal information in Video Compression (LSTVC) and its enhanced version LSTVC+. At first, we design a temporal prior to provide richer mutual information between the overall prior and the current frame content, facilitating the Gaussian parameter prediction. The temporal prior is initialized with the Intra frame (I frame) in a GOP (Group of Pictures) and can be continuously updated during encoding and decoding. Theoretically, the temporal prior includes the global temporal information from the previously decoded frames in the current GOP, so we utilize it to explore the long-range temporal information for both motion and contextual compression. With the high spatial correlations with the current frame, the temporal prior can assist in contextual encoding and decoding for better reconstruction. For short-range counterparts, in other words, temporal information within a limited number of reference frames, we combine the advantages of flow-based and deformable compensation and propose progressive guided motion compensation to achieve robust yet efficient inter coding. We design a hierarchical structure to achieve multi-scale compensation. In each scale, we use optical flow guidance to generate pixel offsets, avoiding inaccurate and overflowing offset prediction. The smaller-scale compensation results will be used to guide the offset prediction of the next scale, improving the interpretability and stability of the entire compensation process. The progressive guiding scheme can stablize training and boost performance steadily. The effectiveness of LSTVC and LSTVC+ is clearly shown as they outperform the current state-of-the-art learned video compression approaches while maintaining high efficiency. They also surpass the commercial codec x265 placebo preset\footnote{Placebo is the slowest setting in x265 which achieves the best quality among ten presets.}. Furthermore, we conduct experiments towards more practical scenarios by setting intra period to 32. Comparing to the standard reference software, our method can surpass HM-16.20 in terms of PSNR and MS-SSIM. We also achieve competitive rate-distortion performance with VTM-11.0 in terms of MS-SSIM. In summary, the contributions of this paper can be summarized as follows:
	\begin{itemize}
		\item We design a temporal prior that can be updated continuously during compression for long-range temporal information exploration. The temporal prior focuses on reducing spatial statistical dependencies by providing additional mutual information between the overall prior and the latent representation. Besides, with the merit of the updating scheme, the temporal prior contains high spatial correlations with the current frame, so it can further boost the reconstruction performance by assisting in contextual compression.
		\item To achieve inter-frame coding, progressive guided motion compensation is proposed for deep mining of short-range temporal information. Specifically, we inherit the stability and flexibility from flow-based and deformable compensation and design a robust hierarchical architecture with a flow-to-kernel and scale-by-scale guiding strategy.
		\item The proposed technologies focus on exploring temporal information and are efficient yet easy to integrate into existing compression frameworks. Embedded by the proposed modules, our method can outperform other state-of-the-art video compression methods by a large margin in terms of both PSNR and MS-SSIM.
	\end{itemize}
	
	The remainder of this paper is organized as follows. Section \uppercase\expandafter{\romannumeral2} introduces the related work. In Section \uppercase\expandafter{\romannumeral3}, we present the LSTVC framework with detailed discussions about the proposed temporal prior and PGMC. In Section \uppercase\expandafter{\romannumeral4}, we show the performance of our proposed framework and discuss the importance of each module, and then we conclude in Section \uppercase\expandafter{\romannumeral5}.
	
	\section{Related Work}
	
	\subsection{Learned Image Compression}
	The pioneer work of Ball\'{e} \emph{et al.} \cite{Balle17a,balle2018variational} modeled the image compression framework as variational auto-encoder and optimized the network in an end-to-end fashion. Besides, for non-linear transform, they introduced generalized divisive normalization (GDN) to Gaussianize the input image into latent representation. Following the progress in probabilistic generative models, Minnen \emph{et al.} \cite{minnen2018joint} proposed autoregressive priors and used masked convolution to reduce spatial statistical redundancy with decoded latents. On top of that, Gaussian Mixture Likelihoods \cite{cheng2020learned} were designed to more accurately model the distribution of the latent representation. As for transform enhancement, Ma \emph{et al.} \cite{ma2020end} proposed a novel wavelet-like transform framework for learned image compression. Gao \emph{et al.} \cite{gao2021neural} used frequency decomposition to decouple the different frequencies and transform them separately. Recent image compression methods \cite{cheng2020learned,ma2020end,gao2021neural} can surpass the intra mode of VTM in terms of both PSNR and MS-SSIM. Except PSNR and MS-SSIM oriented compression, learned image compression methods can also have a greater advantage in targeting some specific problems than traditional codec. Cai \emph{et al.} \cite{8943263} proposed an image compression method which can automatically allocate bits to the region of interest. Mentzer \emph{et al.} \cite{mentzer2020high} explored GAN \cite{creswell2018generative} and perceptual loss \cite{zhang2018unreasonable} to specifically optimize the subjective quality.
	
	\begin{figure*}[t]
		\centering
		\includegraphics[width=0.76\textwidth]{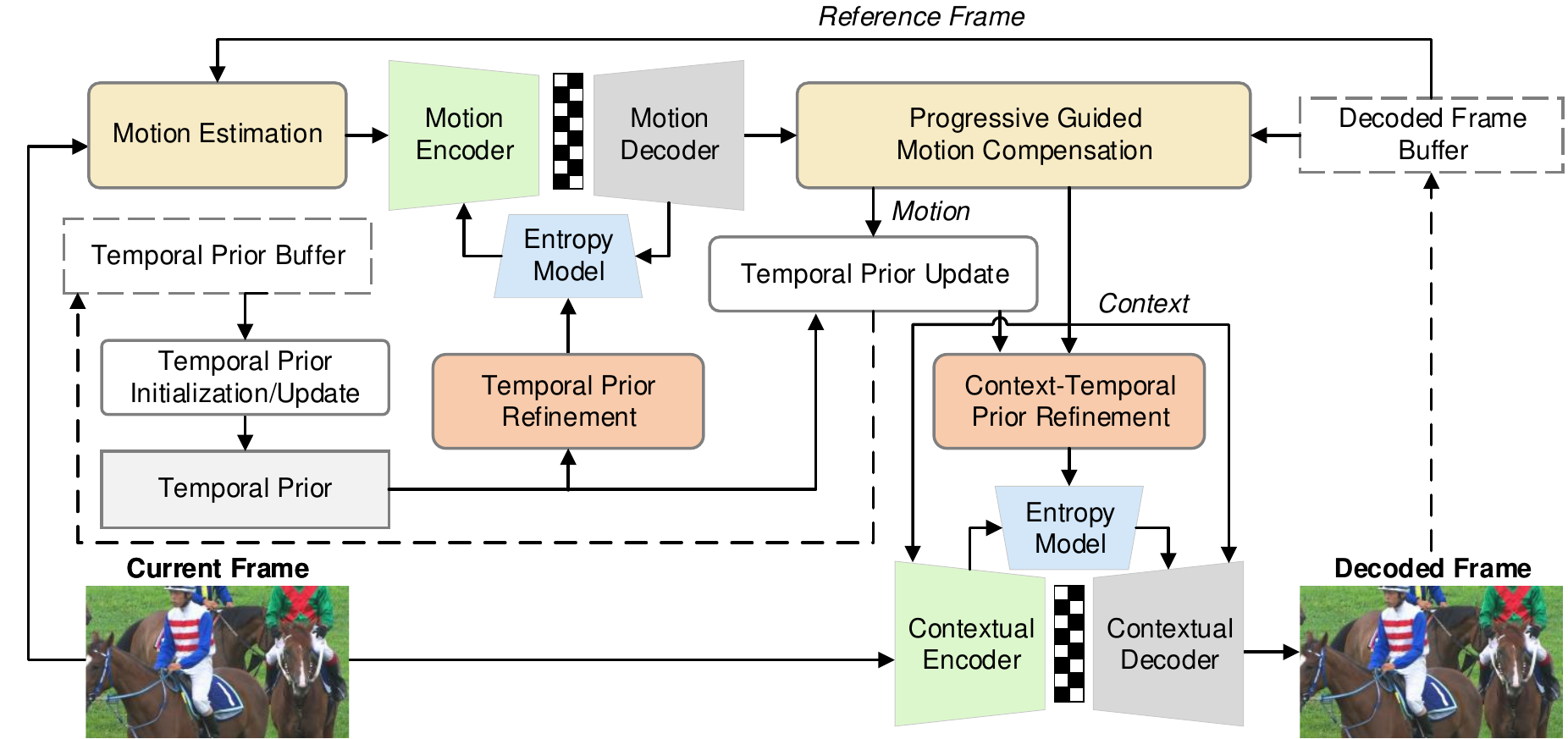}
		\caption{Overview of our proposed video compression framework LSTVC. We extract the temporal information from each decoded frame and supplement the temporal prior with the extracted information. Besides, the temporal prior will be updated explicitly with the decoded motion vectors during compression and should be encoded into the latent representation before participating in the entropy model. The latent representation from motion and contextual compression will be compressed into/decompressed from a bitstream by an arithmetic encoder/decoder. For contextual compression, we use long- and short-range temporal information simultaneously to facilitate distribution parameter prediction.}
		\label{fig:overview}
	\end{figure*}
	
	\subsection{Learned Video Compression}
	In the past few years, learned video compression methods have attracted more attention among researchers. For instance, Wu \emph{et al.} \cite{Wu:2018} creatively treated the video compression task as frame interpolation and achieved comparable performance with H.264 (x264 LDP veryfast) \cite{x264}. However, the framework encoded the motion information by existing non-deep-learning-based coding methods and was unable to be optimized jointly. The Deep Video Compression (DVC) \cite{Lu2019CVPR} was proposed as the first end-to-end optimized video compression algorithm. Unlike the traditional codec framework, DVC replaced motion vectors with optical flow, making joint learning possible. Yang \emph{et al.} \cite{yang2020learning} designed a novel RNN-based compression method HLVC and utilized hierarchical quality layers for compressed video enhancement. Lu \emph{et al.} \cite{lu2020end} extended their DVC with auto-regressive priors and introduced a variable-rate module for building a more flexible framework. Hu \emph{et al.} \cite{hu2020improving} adopted switchable multi-resolution representations for the flow maps based on the RD criterion. Yılmaz \emph{et al.} \cite{9667275} proposed a learned hierarchical bidirectional video codec and surpassed HM-16.23 in terms of MS-SSIM. Hu \emph{et al.} \cite{hu2021fvc} utilized deformable convolution to accomplish compensation in the feature domain, alleviating the errors introduced by simple pixel-to-pixel mapping. Li \emph{et al.} shifted the paradigm from residual coding to conditional coding. They learned temporal contexts from the previously decoded frame and let the model explore the temporal correlation to remove the redundancy.

	\section{Proposed Method}
	In this section, we first introduce the proposed framework and the overall inference process. Then we describe the temporal prior and discuss how it works for compression. Next, we introduce the progressive guided motion compensation in detail. Finally, the training scheme and loss functions are present.
	\subsection{Framework Description}
	We follow the conditional coding-based method DCVC \cite{li2021deep} to build our baseline framework. DCVC abnegates the residue coding-based paradigm and has demonstrated that the contextual information can successfully explore correlation in the video, boosting rate-distortion performance significantly. At first, we formulate a general conditional coding-based compression framework according to the transform coding theory \cite{goyal2001theoretical}. Let $I_{t}$ and $\hat{I}_{t} (t \in {0, 1, ...}) $ denote a sequence of frames and its reconstructed version, and the compression and decompression process of  $i_{th}$ frame can be formulated as:
	\begin{equation}
		\begin{aligned}
			\boldsymbol{m} &=ME(\hat I_{i-1},I_{i}),\\
			\boldsymbol{y^{m}}&=g_a^m(\boldsymbol{m};\boldsymbol{\phi^m}),\\
			\boldsymbol{\hat y^{m}}&=Q(\boldsymbol{y^{m}}),\\
			\boldsymbol{\hat m}&=g_s^m(\boldsymbol{\hat y^{m}};\boldsymbol{\theta^m})\\
			\boldsymbol{cxt} &=MC(\hat I_{t-1},\boldsymbol{\hat m}),\\
			\boldsymbol{y^{c}}&=g_a^c(\boldsymbol{I_i};\boldsymbol{cxt};\boldsymbol{\phi^c}),\\
			\boldsymbol{\hat y^{c}}&=Q(\boldsymbol{y^{c}}),\\
			\boldsymbol{\hat I_i}&=g_s^c(\boldsymbol{\hat y^{c}};\boldsymbol{cxt};\boldsymbol{\theta^c}),
		\end{aligned}
		\label{eq:video_compression}
	\end{equation}
	where $ME$ and $MC$ represent the motion estimation network and the motion compensation operation. $\boldsymbol{m}$, $\boldsymbol{\hat m}$, $\boldsymbol{y^{m}}$ and $\boldsymbol{\hat y^{m}}$ denote motion vector (MV), decoded MV, latent representation of MV and quantized latent representation of MV. $\boldsymbol{cxt}$, $\boldsymbol{y^{c}}$ and $\boldsymbol{\hat y^{c}}$ are context, latent representation of current frame and corresponding quantized latent representation. $g_a^m$, $g_s^m$, $g_a^c$ and $g_s^c$ are analysis and synthesis transforms for motion and context. $\boldsymbol{\phi^m}$, $\boldsymbol{\theta^m}$, $\boldsymbol{\phi^c}$ and $\boldsymbol{\theta^c}$ are learnable parameters of analysis and synthesis transforms for motion and context, respectively. $Q$ represents round-based quantization.
	\begin{algorithm}[t]
		\KwData{Raw video frame data in one GOP}
		\KwResult{\\ \quad \textbf{Encoder}: Bitstream and temporal prior \\ \quad \textbf{Decoder}: Decoded video frame and temporal prior}
		\textbf{Encoding process:}\\
		\For{$I_{i}$ in current GOP}
		{
			\eIf{$i=1$}{
				Compress $I_{i}$ as Intra frame (I frame)\\
			}
			{
				\eIf{$i=2$}{
					Initialize the temporal prior with the I frame\\
				}
				{
					Update the temporal prior with $\hat { I}_{i-1}$\\
				}
				Generate motion $\boldsymbol m$ between $\hat {I}_{i-1}$ and $I_{i}$\\
				Compress/decompress $\boldsymbol m$ into/from the bitstream with the aid of the temporal prior\\
				Generate context $\boldsymbol{cxt}$ with PGMC\\
				Update the temporal prior with the motion $\boldsymbol{\hat m}$\\
				Compress $I_{i}$ into the bitstream with the aid of $\boldsymbol{cxt}$ and the updated temporal prior\\
			}
		}
		\textbf{Decoding process:}\\
		\For{$I_{i}$ in current GOP}
		{
			\eIf{$i=1$}{
				Decompress $\hat{I_{i}}$ from the bitstream\\
			}
			{
				\eIf{$i=2$}{
					Initialize the temporal prior with the I frame\\
				}
				{
					Update the temporal prior with $\hat{I}_{i-1}$\\
				}
				Decompress $\hat{\boldsymbol m}$ from the bitstream with the aid of the temporal prior\\
				Generate context $\boldsymbol{cxt}$ with PGMC\\
				Update the temporal prior with the motion $\hat {\boldsymbol m}$\\
				Decompress $\hat{I}_{i}$ from the bitstream with the aid of $\boldsymbol{cxt}$ and the updated temporal prior\\
			}
		}
		\caption{Inference process of the proposed method.}
		\label{inference process}
	\end{algorithm}
	As for the rate optimization, the framework follows the previous work \cite{Balle17a,minnen2018joint} and models each latent representation element as a Gaussian convolved with a unit uniform distribution. For the description, we use the hyperprior \cite{Balle17a} as side information to predict the mean $\mu$ and scale $\sigma$ of each element of $\boldsymbol y$ to reduce statistic dependencies. The Gaussian parameter prediction process can be formulated as:
	\begin{equation}
		\centering
		\begin{aligned}
			\boldsymbol{z}=&h_a(\boldsymbol{y};\boldsymbol{\phi_h}),\\
			\boldsymbol{\hat z}=&Q(\boldsymbol{z}),\\
			p_{\boldsymbol{\hat y}}(\boldsymbol{\hat y}|\boldsymbol{\hat z},\boldsymbol{\theta_h})=&\prod \limits_{i}(\mathcal N(\mu_i,\sigma_i^2)*\mathcal U(-\frac{1}{2},\frac{1}{2}))(\boldsymbol{\hat y_i})\\
			with& \quad \mu_i, \sigma_i = h_s(\boldsymbol{\hat z};\boldsymbol{\theta_h})
			\label{eq:entropy_model}	
		\end{aligned}
	\end{equation}
	
	where $h_a$, $h_s$, $\phi_h$ and $\theta_h$ are the hyper analysis and synthesis transforms and their learnable parameters in the hyperprior. $\boldsymbol{z}$ and $\boldsymbol{\hat z}$ stand for the side information and its quantized version. As for the conditional coding, the formulation of contextual entropy model goes to:
	\begin{equation}
		\begin{aligned}
			p_{\boldsymbol{\hat y}}(\boldsymbol{\hat y}|\boldsymbol{\hat z},\boldsymbol{cxt},\boldsymbol{\theta_{ef}},\boldsymbol{\theta_h},\boldsymbol{\theta_{ce}})=\prod \limits_{i}(\mathcal N(\mu_i,\sigma_i^2)*\mathcal U(-\frac{1}{2},\frac{1}{2}))(\boldsymbol{\hat y_i})\\
			with \quad \mu_i, \sigma_i = g_{ef}(\phi,\psi;\boldsymbol{\theta_{ef}}),
			\phi=h_s(\boldsymbol{\hat z};\boldsymbol{\theta_h}), \\
			and\quad\psi= g_{ce}(\boldsymbol{cxt};\boldsymbol{\theta_{ce}})
		\end{aligned}
		\label{eq:contextual_entropy_model}	
	\end{equation}
	where $g_{ef}$, $g_{ce}$, $\boldsymbol{\theta_{ef}}$ and $\boldsymbol{\theta_{ce}}$ denote the entropy fusion network (the \emph{Entropy Parameters} network in \cite{minnen2018joint}), the context encoder network and their learnable parameters. $i$ is the index for each element of the latents. $\mathcal N$ and $\mathcal U$ stand for a Gaussian distribution model and a uniform distribution centered on $\boldsymbol{\hat y_i}$, respectively.
	
	In this paper, we extend the framework defined above with enhanced long-range and short-range temporal information and explore their importance in depth for video compression. \textbf{Note that to avoid confusion, we rename the temporal prior in DCVC to context prior since it contains only the motion compensation results of the previous frame features (referred to as short-range temporal information in this paper).} In a word, we name the short-range temporal information from motion compensation as context prior, and refer to our proposed prior containing holistic temporal information among the decoded frames as the temporal prior. 
	
	Fig.~\ref{fig:overview} provides a high-level overview of our video compression model. Specifically, we design a temporal prior that can be continuously updated during the inference process for long-range temporal information utilization. Note that our proposed model is designed for a low-latency scenario, so in one GOP, we initialize the temporal prior with the I frame before compressing the first Predictive frame (P frame). During P frame compression, the motion and contextual compression process are both conditioned on the temporal prior. Besides, we update the proposed prior after obtaining the decoded motion to generate prior with a stronger spatial correlation with the original frame, facilitating the entropy model to develop more accurate Gaussian parameters. To better utilize short-range temporal information and achieve robust and effective compensation, we propose progressive guided motion compensation (PGMC), which will be described in Section \ref{sec:PGMC}. Algorithm \ref{inference process} presents the detailed inference process, including encoding and decoding. Note that we use the high-efficient multistage context model in \cite{lu2022high} to replace auto-regressive prior for avoiding parallel-unfriendly operations. We also provide two versions of the framework LSTVC and LSTVC+, the latter with a more complex transformation network. The detailed description about these two architectures are provided in the open-source codes\footnote{\href{https://github.com/Huairui/LSTVC}{https://github.com/Huairui/LSTVC}}.
	
	\subsection{Temporal Prior} \label{sec:temporal Prior}
	In learned image and video compression, priors of the entropy model play a significant role in predicting more accurate parameters of Gaussian distribution and enhancing rate-distortion performance. Valuable prior information can greatly reduce the Shannon cross entropy between the predicted distribution of the latent representation and the actual marginal distribution. For instance, Ball\'{e} \emph{et al.} \cite{balle2018variational} proposed a hyperprior to enhance non-parametric factorized density model, predicting scale parameters using side information. Based on this, Minnen \emph{et al.} \cite{minnen2018joint} proposed joint autoregressive and hierarchical priors to predict the distribution parameters of the current latent from the decoded latents by masked convolution, reducing spatial statistic dependencies successfully. 
	
	Since hyperprior \cite{balle2018variational} and auto-regressive prior \cite{minnen2018joint} have achieved promising performance in image compression, many video compression methods \cite{Lu2019CVPR,lu2020end,lin2020m,hu2020improving,hu2021fvc} directly follow the image compression work and adopt hyperprior or autoregressive context model for distribution parameter prediction for compression. For instance, DVC and DVC\_Pro utilized hyperprior and auto-regressive prior to compress motion and residual, respectively. M-LVC \cite{lin2020m} directly used fully-factorized entropy model. Agustsson \cite{agustsson2020scale} chose the GMM model to predict mean and scale parameters of Gaussian distribution without auto-regressive operation. 
	
	However, video content has its own unique characteristics: high spatial redundancy in the temporal dimension, but few methods focus on this and apply it to the prior generation. The most related work is NVC \cite{liu2020learned} which proposed joint spatial-temporal prior. By using convLSTM \cite{shi2015convolutional}, NVC updated the temporal prior with the corresponding latents of the previous frame. From their ablation study (refer to Fig.~9 in \cite{liu2020learned}), we can infer that there is still much room to improve the ability in statistical dependencies removing  by temporal information. In this paper, we propose a more intuitive yet effective prior updating and usage scheme and deeply mine the importance of long-range temporal information.
	
	\subsubsection{Initialization and Updating}
	We initialize the temporal prior for each GOP before compressing the first P frame, and we extract prior features from the I frame using three cascaded convolutional layers for initialization. The sizes of temporal prior buffer are set to $(64, H/4, W/4)$, where $(3, H, W)$ denotes the sizes of the input frames. Besides, we update the proposed temporal prior under two conditions. First, to generate a temporal prior with a higher spatial correlation to the current frame, we align and update the temporal prior with the decoded motion vector before contextual compression. Second, except for the first P frame, we update the prior with the decoded frame contents before compressing the current motion vectors. At the first updating, we supplement the temporal with the reference frame. This updating scheme makes the prior tend to store information close to the current frame and discard information that is too far from the current frame. With the merit of the first updating scheme, the temporal prior can dynamically preserve and discard long-range temporal information. 
	
	\subsubsection{Prior for Motion Compression}
	To achieve high efficient motion estimation, we use SPyNet \cite{ranjan2017optical} to generate optical flow as the motion vector, which contains pixel-level inter-frame motion information. The alignment of key points, as well as the structural information are core information for motion compensation. Since the temporal prior has rich contour and texture information from holistic decoded frames within the current GOP, it can facilitate Gaussian parameter prediction for the motion compression. 
	
	Although the proposed prior encompass much temporal information from previously decoded frames, it is counter-intuitive to directly feed the temporal prior into the entropy model. Inspired by DCVC, we design a prior encoder to transform the temporal prior into the motion-correlated latent representation. The formulation of motion entropy model can be stated as:
	\begin{equation}
		\centering
		\begin{aligned}
			p_{\boldsymbol{\hat y}}(\boldsymbol{\hat y}|\boldsymbol{\hat z},\boldsymbol{tpr},\boldsymbol{\theta_{ef}},\boldsymbol{\theta_h},\boldsymbol{\theta_{pe}})=\prod \limits_{i}(\mathcal N(\mu_i,\sigma_i^2)*\mathcal U(-\frac{1}{2},\frac{1}{2}))(\boldsymbol{\hat y_i})\\
			with \quad \mu_i, \sigma_i = g_{ef}(\phi,\psi;\boldsymbol{\theta_{ef}}),
			\phi=h_s(\boldsymbol{\hat z};\boldsymbol{\theta_h}), \\
			and\quad\psi= g_{pe}(\boldsymbol{tpr};\boldsymbol{\theta_{pe}})
			\label{eq:motion_entropy_model_wTP}	
		\end{aligned}
	\end{equation}
	where $\boldsymbol{tpr}$ denotes the temporal prior, $g_{pe}$ and $\boldsymbol{\theta_{pe}}$ are the prior encoder and its learnable parameters. 
	\begin{figure*}[t]
		\centering
		\includegraphics[width=0.84\textwidth]{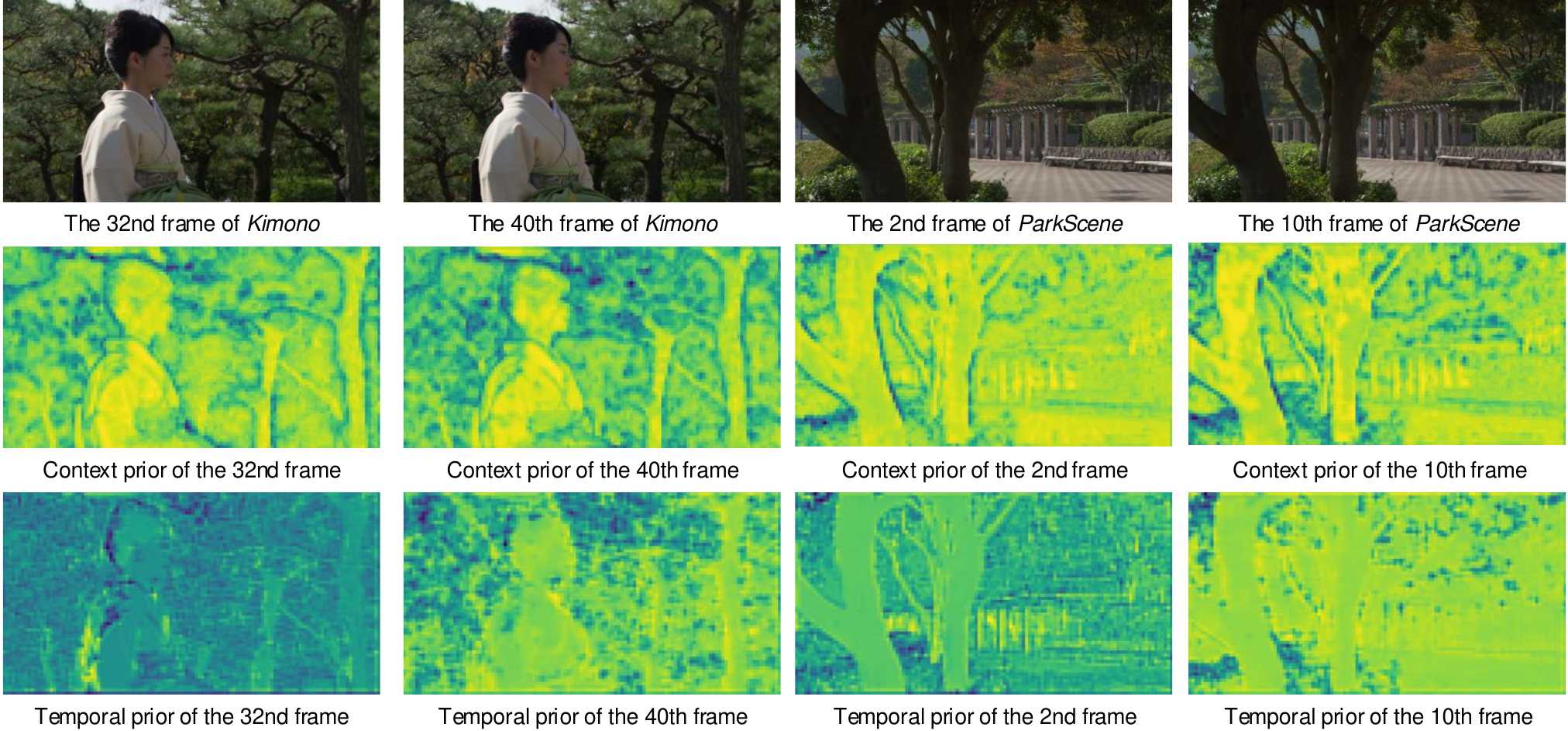}
		\caption{Visualization of the refine context prior and temporal prior.}
		\label{fig:vis_prior}
		\vspace{-10pt}
	\end{figure*}
	\subsubsection{Prior for Contextual Compression}
	In the conditional coding-based framework, the whole contextual compression should be conditioned on the context prior $\boldsymbol{cxt}$ as described in the formulation (\ref{eq:video_compression}) and (\ref{eq:contextual_entropy_model}). In our compression scheme, we update the temporal prior with the decoded motion vector and ensure that the temporal prior has strong spatial correlations with the current frame. Then the contextual compression model should be conditioned on our temporal prior either.

	As we mentioned above, the updated temporal prior has strong spatial correlations with the current frame and can significantly reduce spatial statistical dependencies. In other words, without the high computational complexity and parallel-unfriendly entropy model (such as the auto-regressive context used in \cite{lu2020end,yang2020learning,li2021deep}), our framework can still achieve superior RD performance with the aid of the temporal prior. For the contextual entropy model, we design a hybrid prior encoder to transform the temporal prior and the context prior into the latent representation. The contextual entropy model can be formulated as:
	\begin{equation}
		\centering
		\begin{aligned}
			p_{\boldsymbol{\hat y}}(\boldsymbol{\hat y}|\boldsymbol{\hat z},\boldsymbol{cxt},\boldsymbol{tpr},\boldsymbol{\theta_{ef}},\boldsymbol{\theta_h},\boldsymbol{\theta_{hpe}})= \qquad\qquad\qquad\qquad
			\\\prod \limits_{i}(\mathcal N(\mu_i,\sigma_i^2)*\mathcal U(-\frac{1}{2},\frac{1}{2}))(\boldsymbol{\hat y_i})\\
			with \quad \mu_i, \sigma_i = g_{ef}(\phi,\psi;\boldsymbol{\theta_{ef}}),
			\phi=h_s(\boldsymbol{\hat z};\boldsymbol{\theta_h}), \\
			and\quad\psi= g_{hpe}(\boldsymbol{cxt};\boldsymbol{tpr};\boldsymbol{\theta_{hpe}})
		\end{aligned}
		\label{eq:context_entropy_model_wTP}
	\end{equation}
	where $g_{hpe}$ and $\boldsymbol{\theta_{hpe}}$ denote the hybrid context-temporal prior encoder and its learnable parameters.

	\subsubsection{Discussion}
	We notice that for long-range temporal information exploration, the most related work is NVC \cite{liu2020learned} which also updates their temporal prior at a frame recurrent fashion. NVC implicitly uses ConvLSTM and the motion latent representation to update the prior, and the temporal prior is only engaged in the motion entropy model. Regarding design philosophy, our approach differs from NVC in three major ways, which are the core innovations in exploring long-range temporal information. We briefly summarize the proposed	scheme as follows:
	
	\begin{itemize}
	\item \textbf{Initialization:} We initialize the temporal prior with the feature extracted from the I frame. Besides, before feeding the temporal prior into the entropy model, we use the motion and contextual prior encoders to transform the prior into the corresponding latent representation.
	\item \textbf{Updating:} We update the temporal prior under two conditions. First, we re-extract the feature from the previous frame before motion compression, with which we supplement the prior. Furthermore, before contextual compression, we update the prior explicitly with the decoded motion vector, providing the temporal prior with a stronger spatial correlation to the current frame.
	\item \textbf{Usage:} In our framework, the temporal prior is utilized in both motion and contextual entropy models, mainly assisting in removing spatial statistical dependencies. Meanwhile, extra prior encoders are proposed to transform the prior into the corresponding latent representation to avoid the counter-intuitive use of the temporal prior.
	\end{itemize}

	Furthermore, traditional codecs use a preset reference frame number to capture long-range temporal information. However, they cannot choose or update the temporal information dynamically according to the video content. In that case, introducing too much temporal information far from the current frame may cause performance loss. As for our method, the temporal prior will bed supplemented with the reference frame information at the first updating. This updating scheme makes the temporal prior tend to store information close to the current frame and discard information that is too far from the current frame. With the merit of the updating scheme, the temporal prior can dynamically preserve and discard long-range temporal information, which helps avoid bad cases as in traditional codecs.

	Fig.~\ref{fig:vis_prior} shows the visualization results of the refined context and temporal prior. In our implementation, the refined priors are directly engaged in the Gaussian parameter estimation of the contextual entropy model. From the visualization results, we observe two phenomena. One is that the information volume of the temporal prior accumulates during coding, while the feature response of the context prior will not change significantly within a GOP. This is reasonable because information on the temporal prior will be continuously supplemented and updated in the coding progress. Context prior always contains only information from the aligned reference frame, without information accumulation.
		\begin{figure*}[t]
		\centering
		\includegraphics[width=0.83\textwidth]{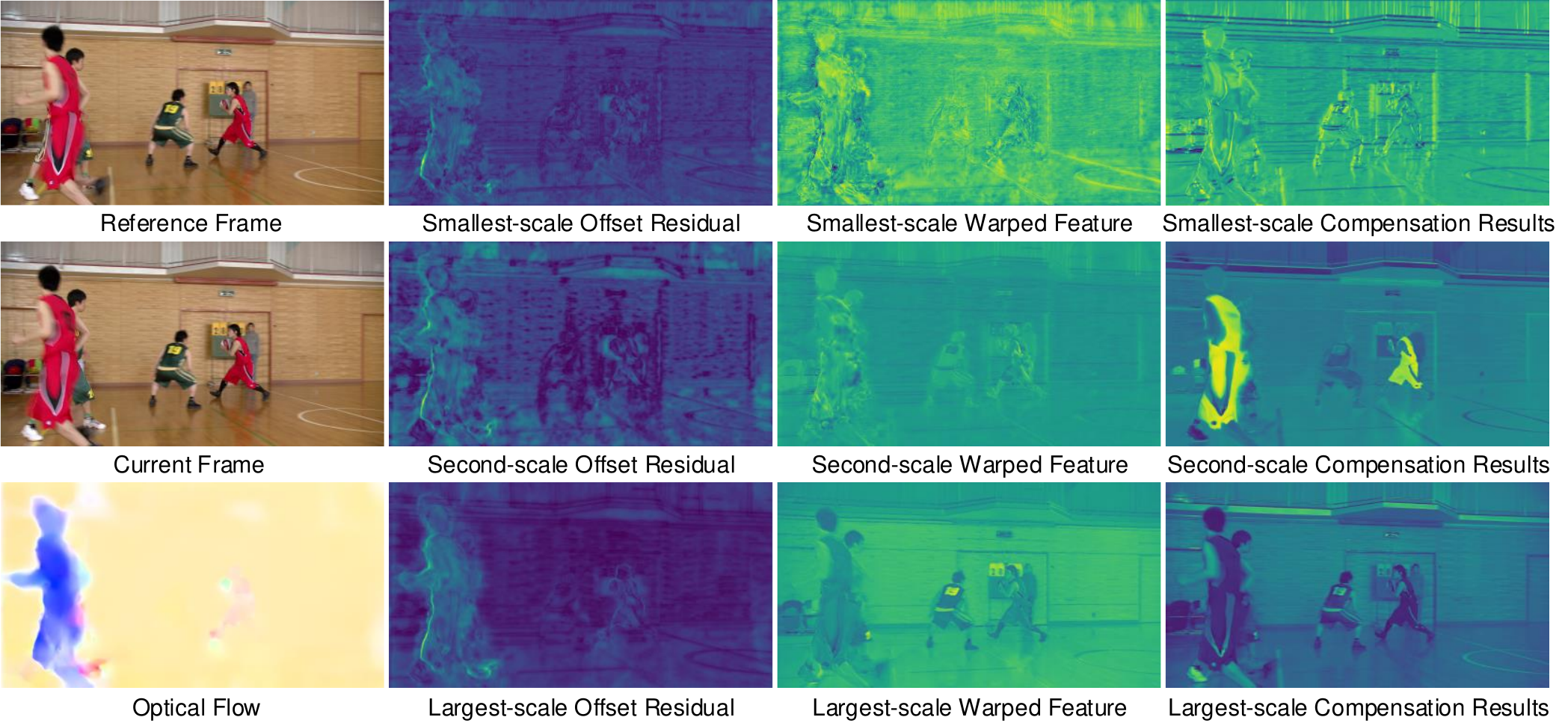}
		\caption{Visualization of the motion vector, offset residual, warped feature and compensation results from PGMC.}
		\label{fig:vis_motion}
	\end{figure*}
	\begin{figure}[t]
		\centering
		\includegraphics[width=0.45\textwidth]{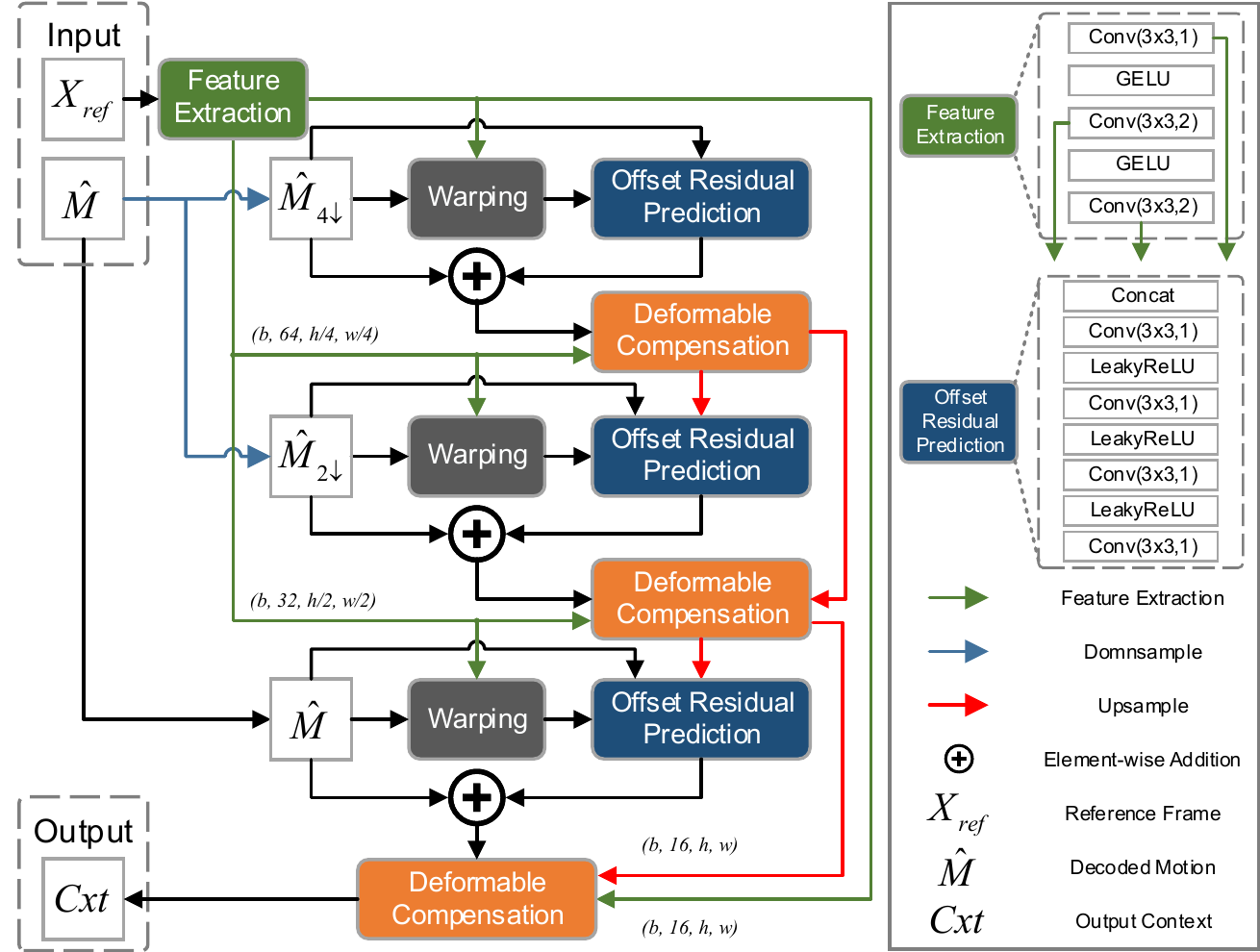}
		\caption{The illustration of the progressive guided motion compensation. The channel number of feature in each scale decreases scale by scale to reduce operation's complexity.}
		\label{fig:pgmc}
	\end{figure}
	
	The second phenomenon is that the feature responses on the temporal prior and context prior are focused on the area with concrete objects, such as humans and trees. In the meantime, the focus areas of these two priors are complementary. For example, the context prior focuses on expressing the main part of the person, and the temporal prior focuses on expressing the marginal part of the person, such as the back and chest. This complementary feature expression also demonstrates that the temporal prior can provide valuable information for regions not attended to by the contextual prior.

	\subsection{Progressive Guided Motion Compensation}\label{sec:PGMC}
	Previous learned compression works have explored short-range temporal information and proposed many effective motion compensation operations. DVC utilized a flow prediction network \cite{ranjan2017optical} to achieve jointly optimization of motion estimation, compensation and compression framework. Liu \emph{et al.} \cite{liu2020learned} designed a one-stage unsupervised approach to estimate implicit flow and facilitate motion estimation. Agustsson \cite{agustsson2020scale} proposed a scale parameter for optical flow and achieved the much more robust compensation performance. Hu \emph{et al.} \cite{hu2021fvc} utilized deformable convolution to accomplish compensation in the feature domain, alleviating the align errors introduced by inaccurate pixel-level operations. In the meantime, from the related work \cite{chan2021understanding} in other video processing tasks, we can conclude that optical flow warping can be seen as a particular case of deformable convolution when the kernel size is set to 1. So there is a consensus \cite{hu2021fvc,wang2022learned} that deformable convolution has higher flexibility and better compensation performance than flow warping. However, high computation burden and unstable training are two remaining problems that deformable compensation will cause \cite{wang2022learned}. 
	
	In this paper, we propose a progressive guided motion compensation (PGMC) scheme which uses the core idea of optical flow guiding convolution and scale-by-scale guided reconstruction and Fig.~\ref{fig:pgmc} shows the details of the proposed PGMC. Without prompt constrain and guidance, kernel offsets are difficult to estimate accurately, and the large range of motion may cause offsets prediction overflow. Inspired by the second-order flow-guided deformable alignment in \cite{Chan_2022_CVPR}, we simplify it to first-order alignment in order to accommodate the task. We use optical flow to guide the generation of the kernel offsets for achieving stable training and accurate compensation. 
	
	Furthermore, to avoid overflow of the offsets prediction, we prioritize motion compensation in the smaller-scale level and guide offset prediction and feature reconstruction for the next scale. Specifically, compensation on the smaller scale is much easier to accomplish since the absolute motion magnitude decreases as the resolution decreases, so smaller-scale compensation aims at coarse offset prediction and frame reconstruction. The offsets generated by guidance participate in motion compensation, and the compensation results continue to guide the offsets prediction for the next scale, forming a progressively guided compensation scheme. Furthermore, the robust scheme allows for lightweight design, so we cut the channel number in larger scale to reduce model complexity. 
	
	Fig.~\ref{fig:vis_motion} shows the visualization of motion vector, offset residual, warped feature and compensation results. It can be observed that the decoded optical flow information is not accurate enough to achieve effective compensation. Still, the offset residual allows for the correction of inaccurate optical flow. On a smaller scale, the offset residual mainly assists in capturing the changes in background information brought about by the camera motion. On the largest scale, the offset residual mainly contributes to processing large and complex movements (such as running athletes). We can infer that the flow-to-kernel guiding strategy reduces the demand for high-precision optical flow in the compensation module, and the PGMC can achieve satisfying compensation with lower entropy of motion information. As for the warped features and compensation results visualization, these two kinds of features are mainly composed of profile and structure information of the current frame in the smallest scale. In the second scale, the features tend to contain object information. The features of the largest scale include complete video frames information, such as background, object outline, and high-frequency details. Different information contained in each scale demonstrates that the compensation process is carried out in an easy-to-hard manner, which also justifies the rationality of the PGMC.

	\subsection{Loss Function}\label{sec:loss}
	DCVC does not add supervision on the context \cite{li2021deep} and wishes to learn the condition in the feature domain automatically. From its ablation study and our experiments, we infer that context prior has a rich spatial correlation with the current frame and can reduce spatial statistical dependencies as auto-regressive prior does. To this end, we train motion-related modules at first to further improve spatial relevance between the context prior and the current frame, plus stabilize training. Specifically, following the training strategy in \cite{wang2022learned}, our framework focus on stabilizing compensation performance at first and only motion-related modules are engaged in training. The rate-compensation loss $L_{r-c}$ is used in this stage:
	\begin{equation}
		L_{r-c}=\lambda \cdot D_{com}+R=\lambda \cdot d(\boldsymbol{cxt},F_{t})+R_{\tilde{m}},
		\label{equa:MCC_loss}
	\end{equation}
	in which $\boldsymbol{cxt}$ denotes the compensation results, $F_{t}$ is the feature extracted from the current frame, and $\lambda$ is the Lagrange multiplier that determines the trade-off between the predicted entropy results $ R $ and the compensation distortion $D_{com}$. 
	
	Our video compression framework try to tradeoff the number of bits used for encoding each frame and reducing distortion between the original input frame $I$ and the reconstructed frame $\hat{I}$. I the end, we use the total RD optimization loss function:
	\begin{equation}
		L_{total}=\lambda \cdot D+R=\lambda \cdot d(\hat{I}_{t},I_{t})+R_{\tilde{m}}+R_{\tilde{c}}
		\label{equa:total_loss}
	\end{equation}
	where $d(I_{t},\hat{I}_{t})$ denotes the distortion between $I_{t} $ and $\hat{I}_{t} $. $R_{\tilde{m}}$ and $R_{\tilde{c}}$ stand for the estimated rate for motion and contextual compression. To train our PSNR model, we use MSE as the distortion function, and as to the MS-SSIM (multi-scale structural similarity) model, we use $d(x, y)=1-$ MS-SSIM$(x,y)$ as the distortion function.
	\begin{figure*}[t]
		\centering
		\begin{subfigure}{.28\textwidth}
			\centering
			\includegraphics[width=\textwidth]{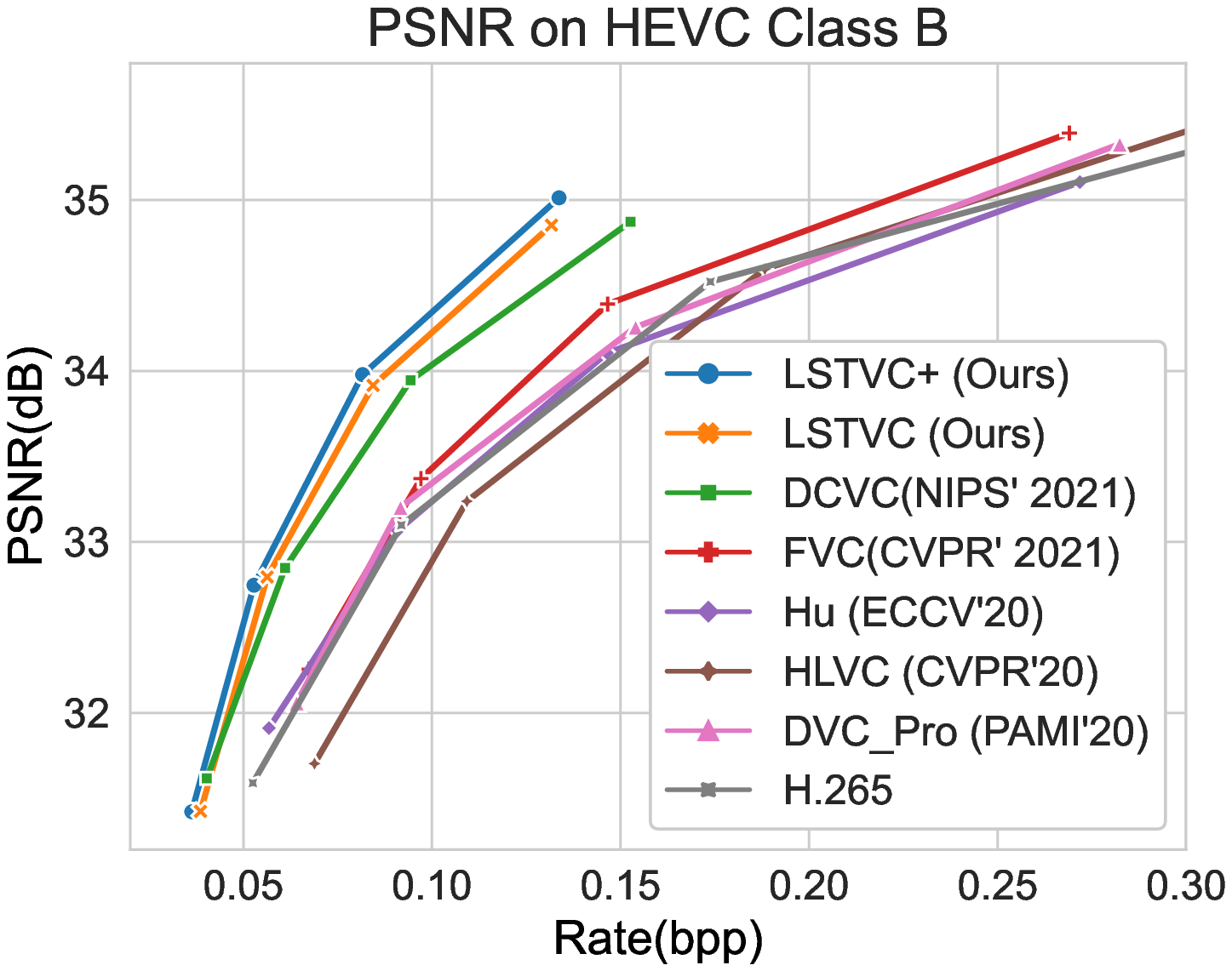}
			\label{ClassB_PSNR}
		\end{subfigure}
		\begin{subfigure}{.28\textwidth}
			\centering
			\includegraphics[width=\textwidth]{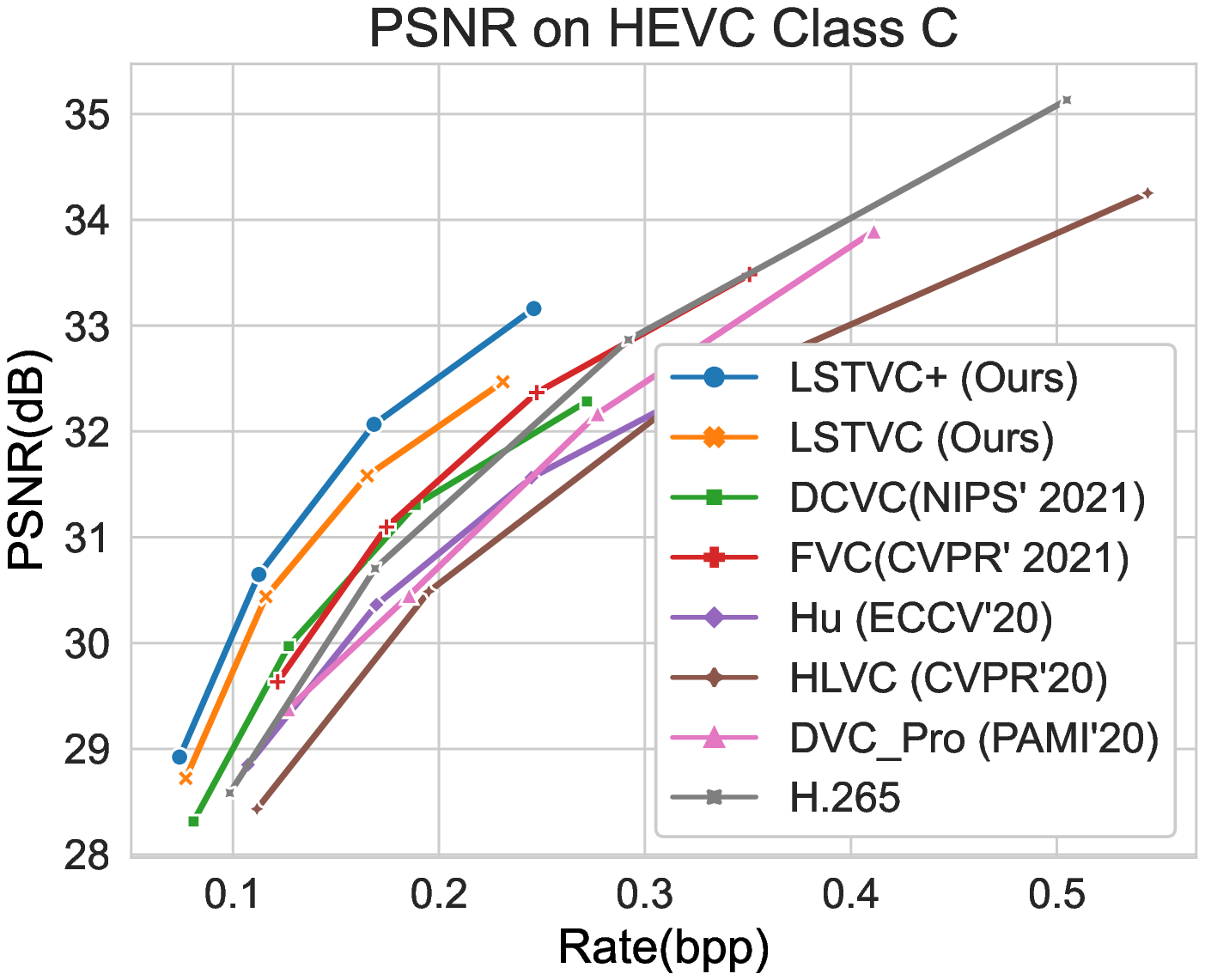}
			\label{ClassC_PSNR}
		\end{subfigure}
		\begin{subfigure}{.28\textwidth}
			\centering
			\includegraphics[width=\textwidth]{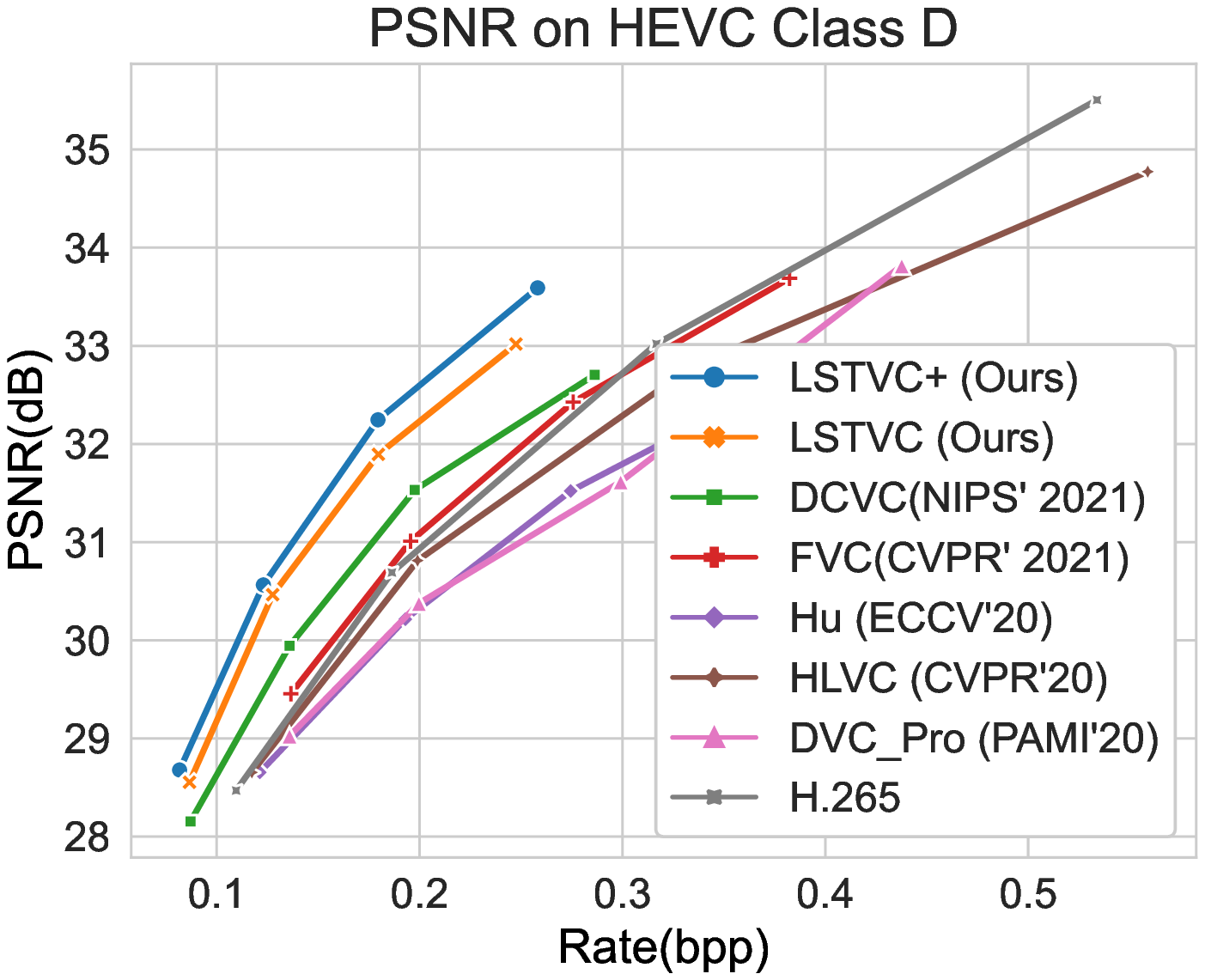}
			\label{ClassD_PSNR}
		\end{subfigure}
		\\
		\begin{subfigure}{.28\textwidth}
			\centering
			\includegraphics[width=\textwidth]{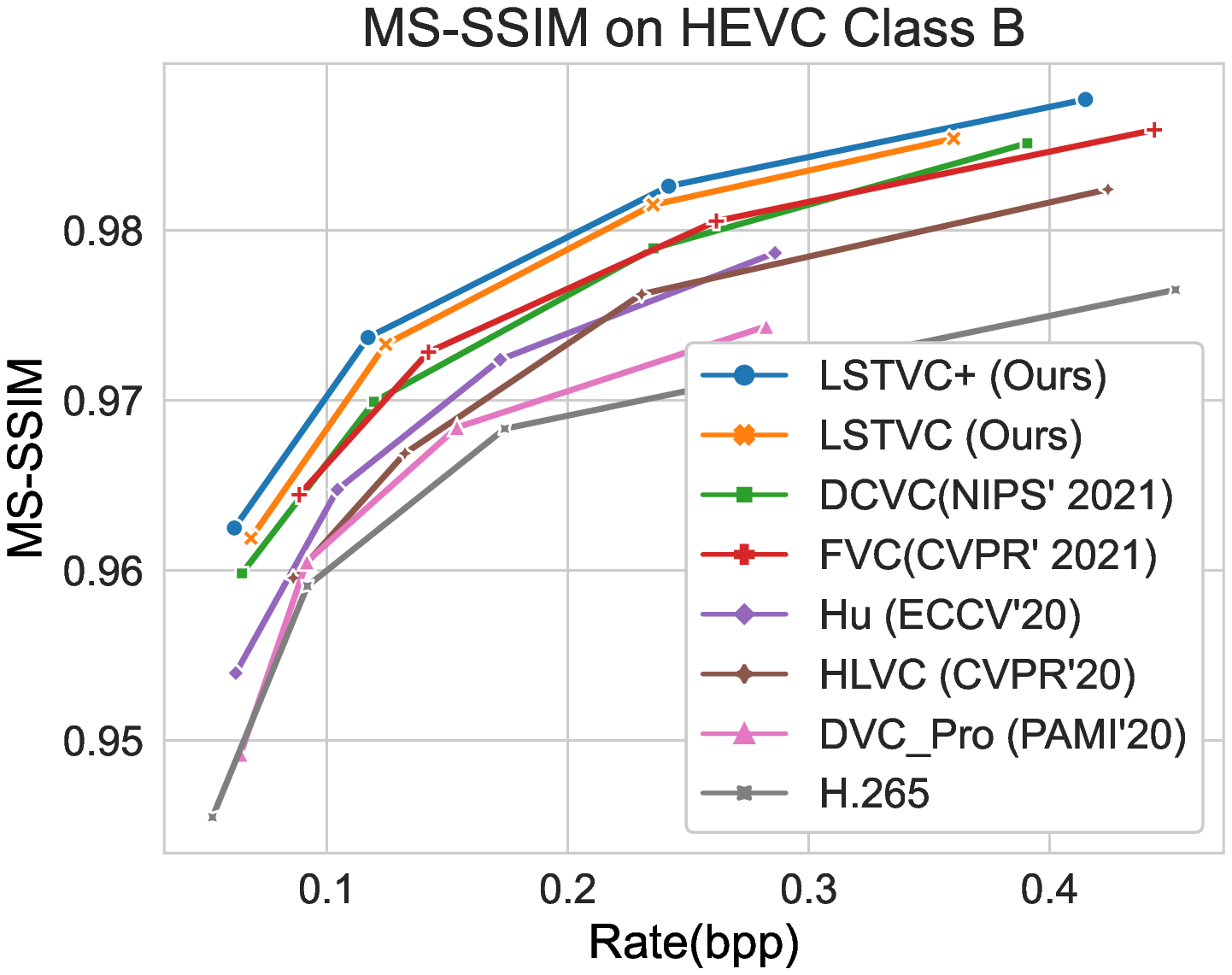}
			\label{ClassB_MSSSIM}
		\end{subfigure}
		\begin{subfigure}{.28\textwidth}
			\centering
			\includegraphics[width=\textwidth]{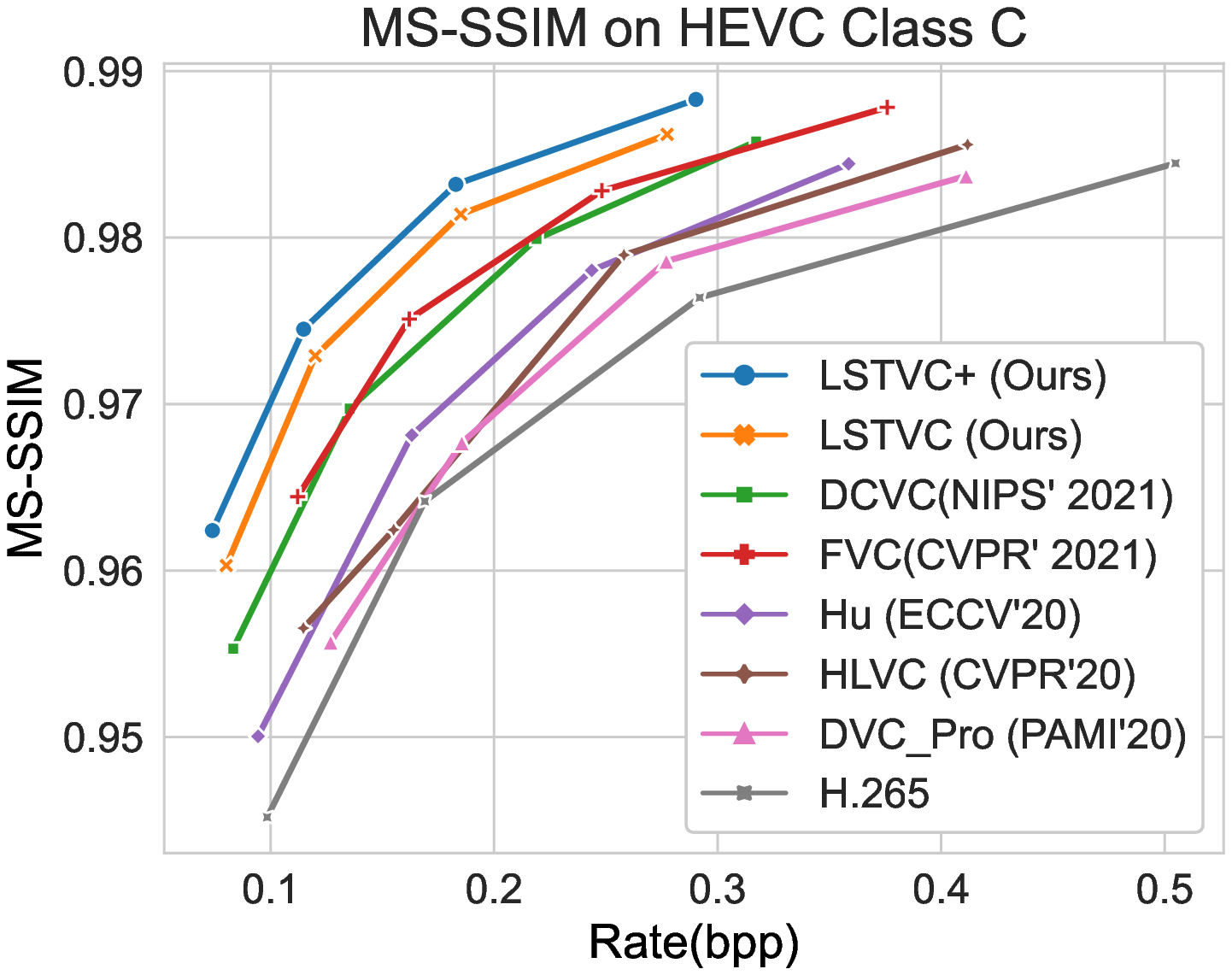}
			\label{ClassC_MSSSIM}
		\end{subfigure}
		\begin{subfigure}{.28\textwidth}
			\centering
			\includegraphics[width=\textwidth]{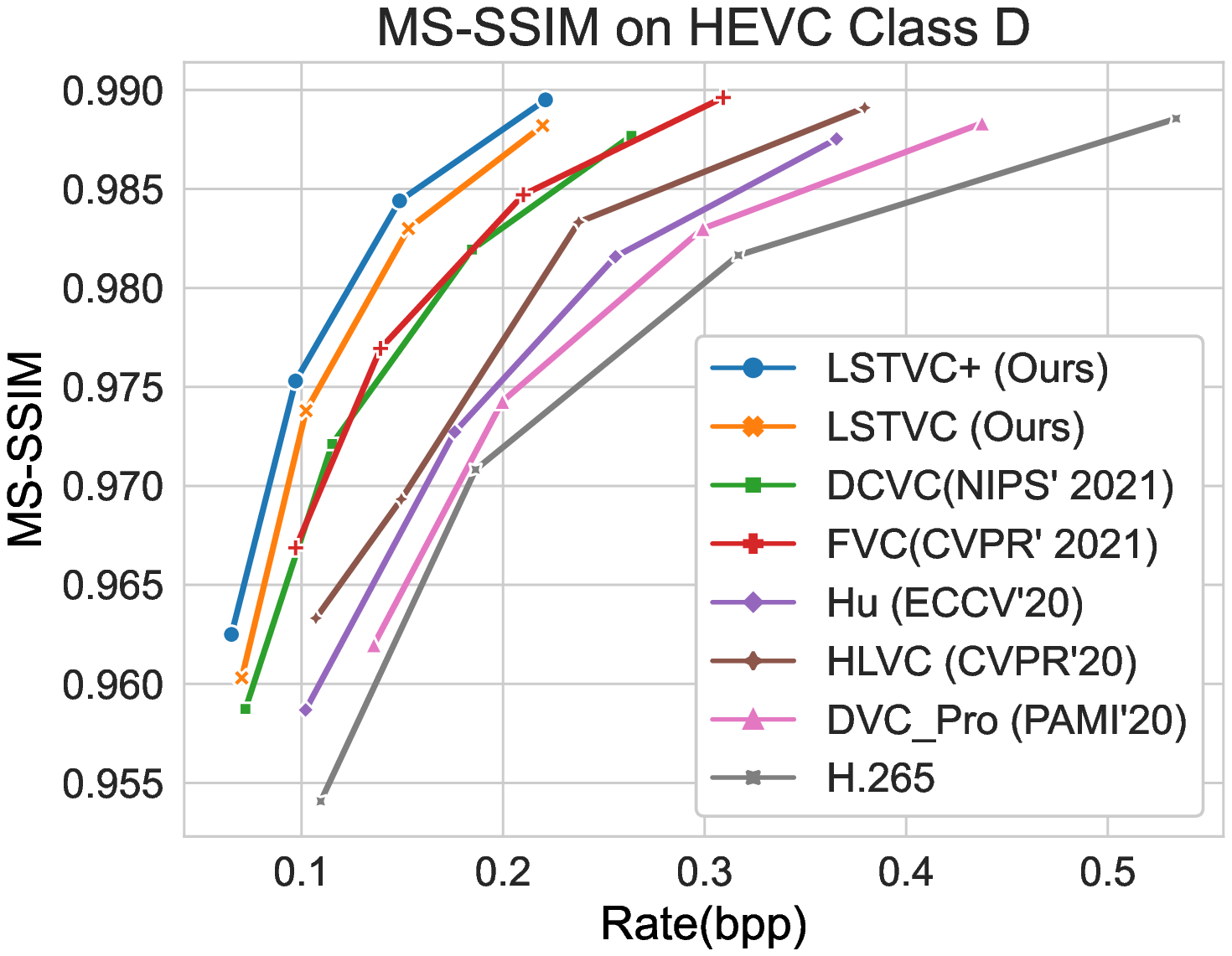}
			\label{ClassD_MSSSIM}
		\end{subfigure}
		\\
		\begin{subfigure}{.28\textwidth}
			\centering
			\includegraphics[width=\textwidth]{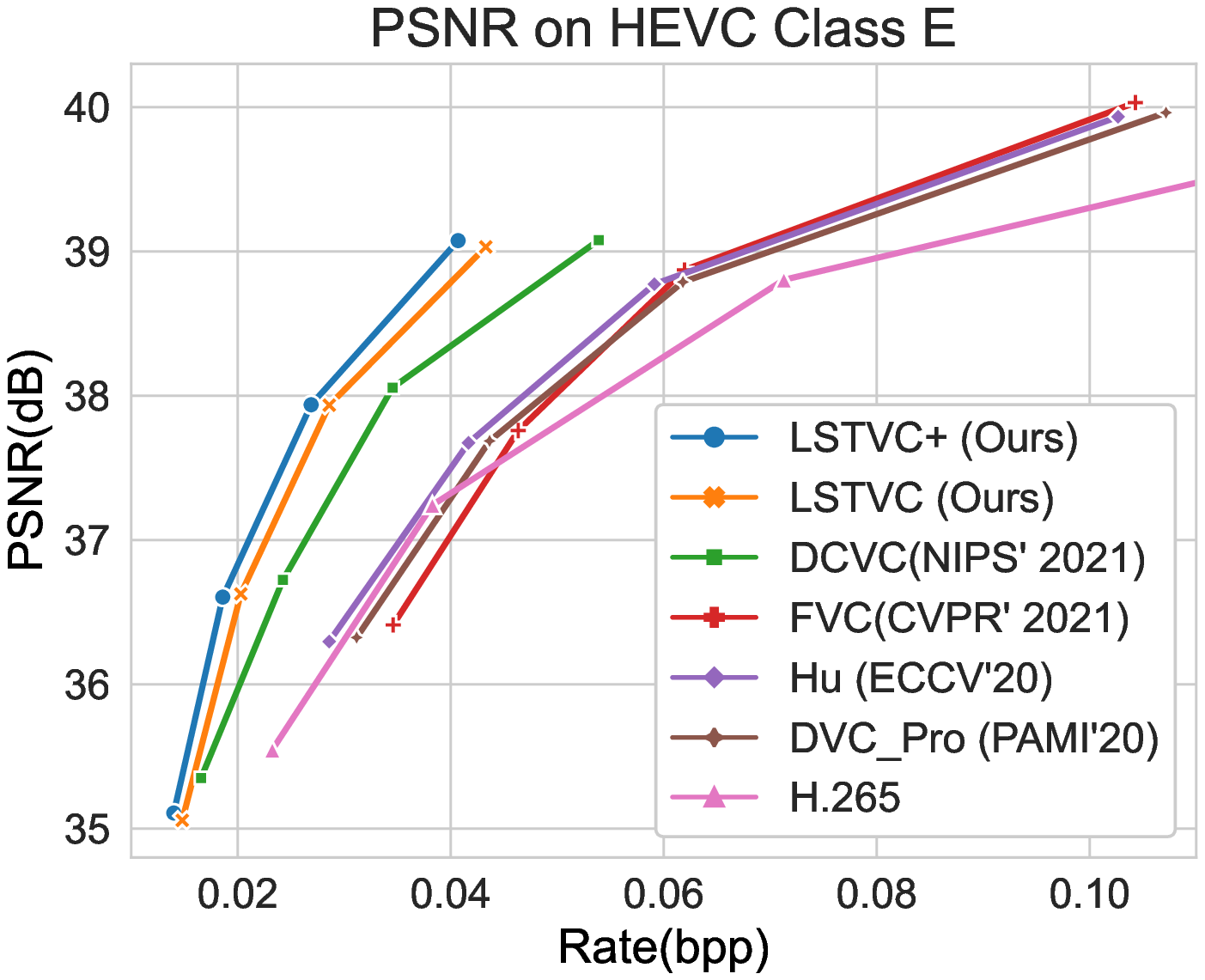}
			\label{ClassE_PSNR}
		\end{subfigure}
		\begin{subfigure}{.28\textwidth}
			\centering
			\includegraphics[width=\textwidth]{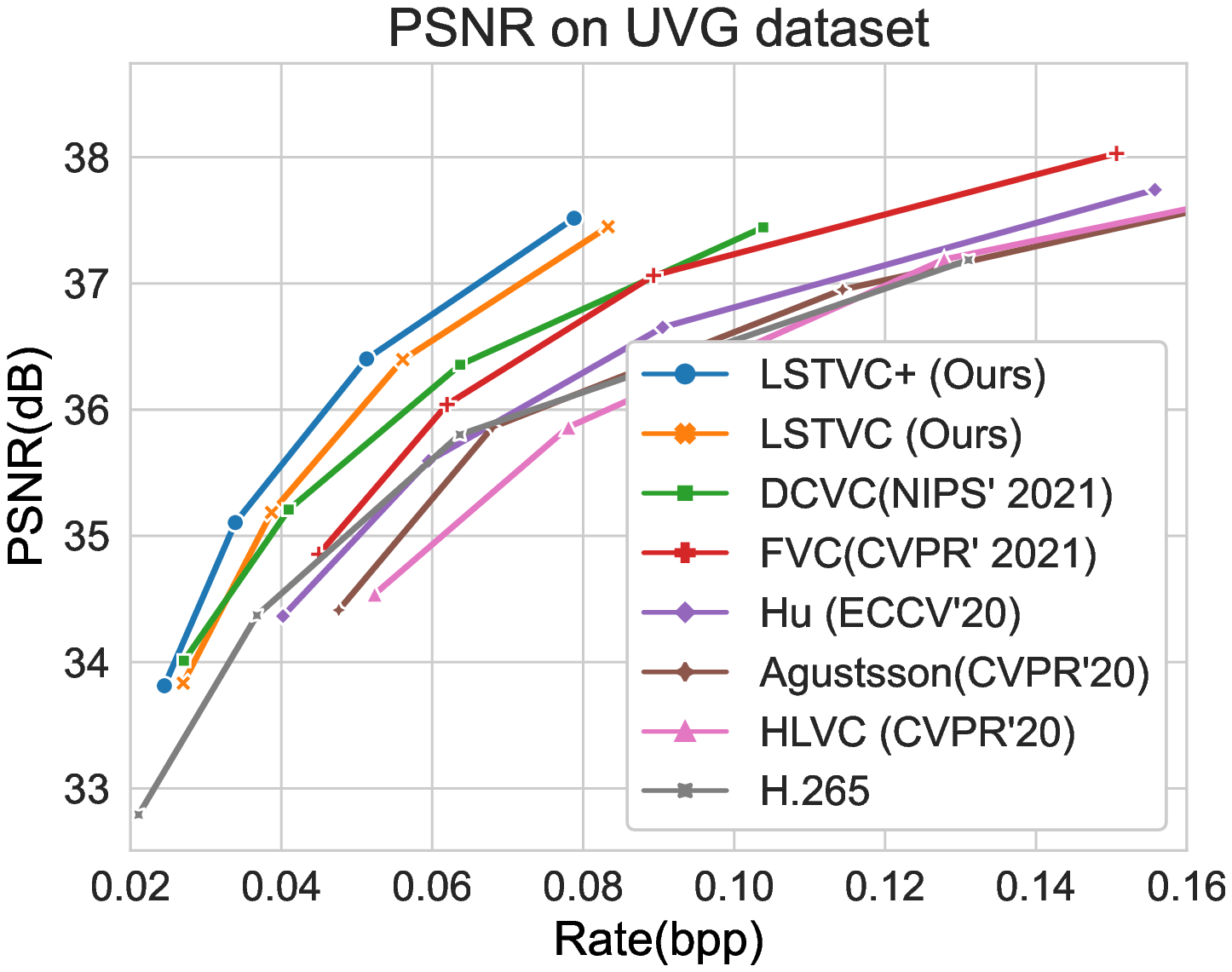}
			\label{UVG_PSNR}
		\end{subfigure}
		\begin{subfigure}{.28\textwidth}
			\centering
			\includegraphics[width=\textwidth]{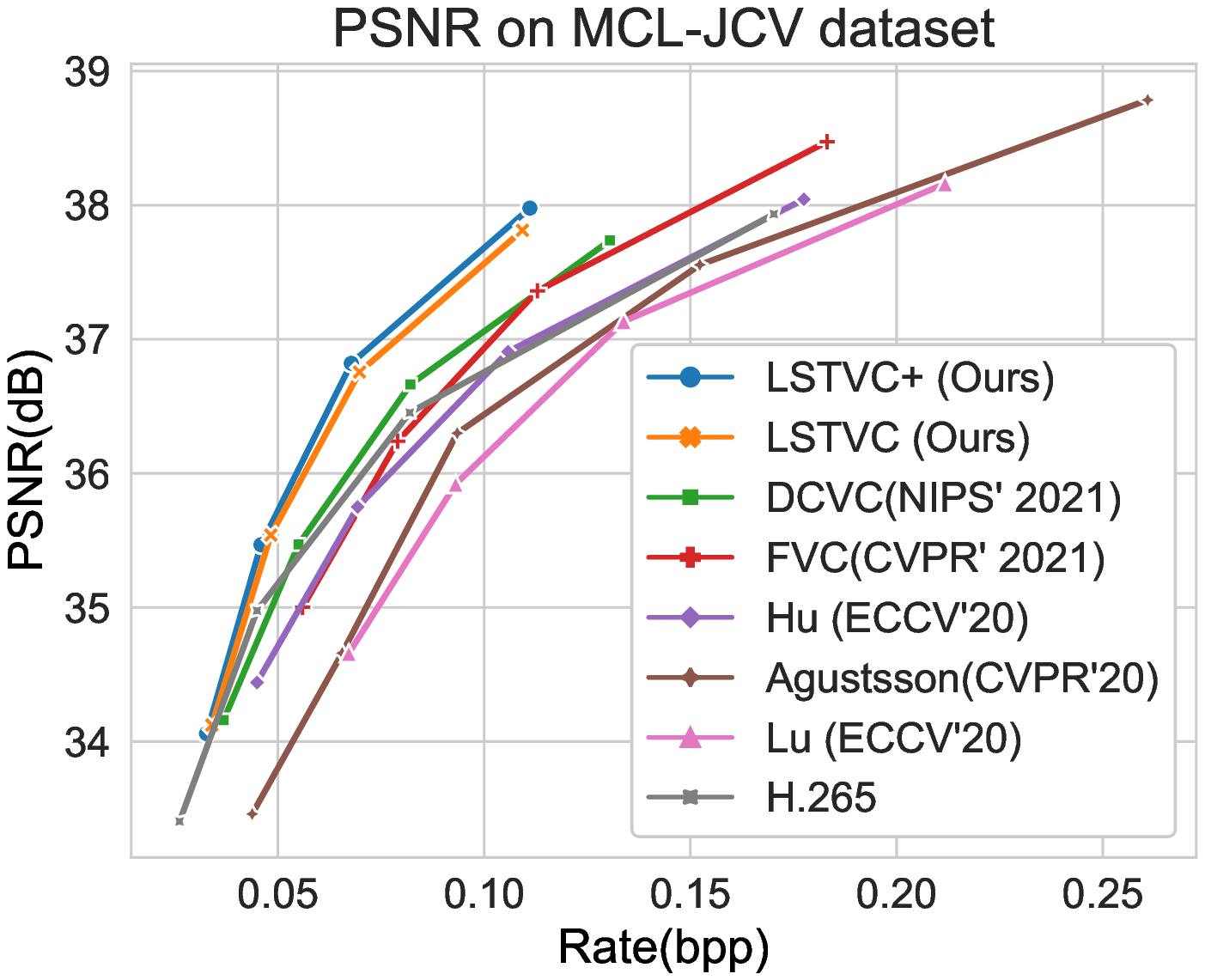}
			\label{MCLJCV_PSNR}
		\end{subfigure}
		\\
		\begin{subfigure}{.28\textwidth}
			\centering
			\includegraphics[width=\textwidth]{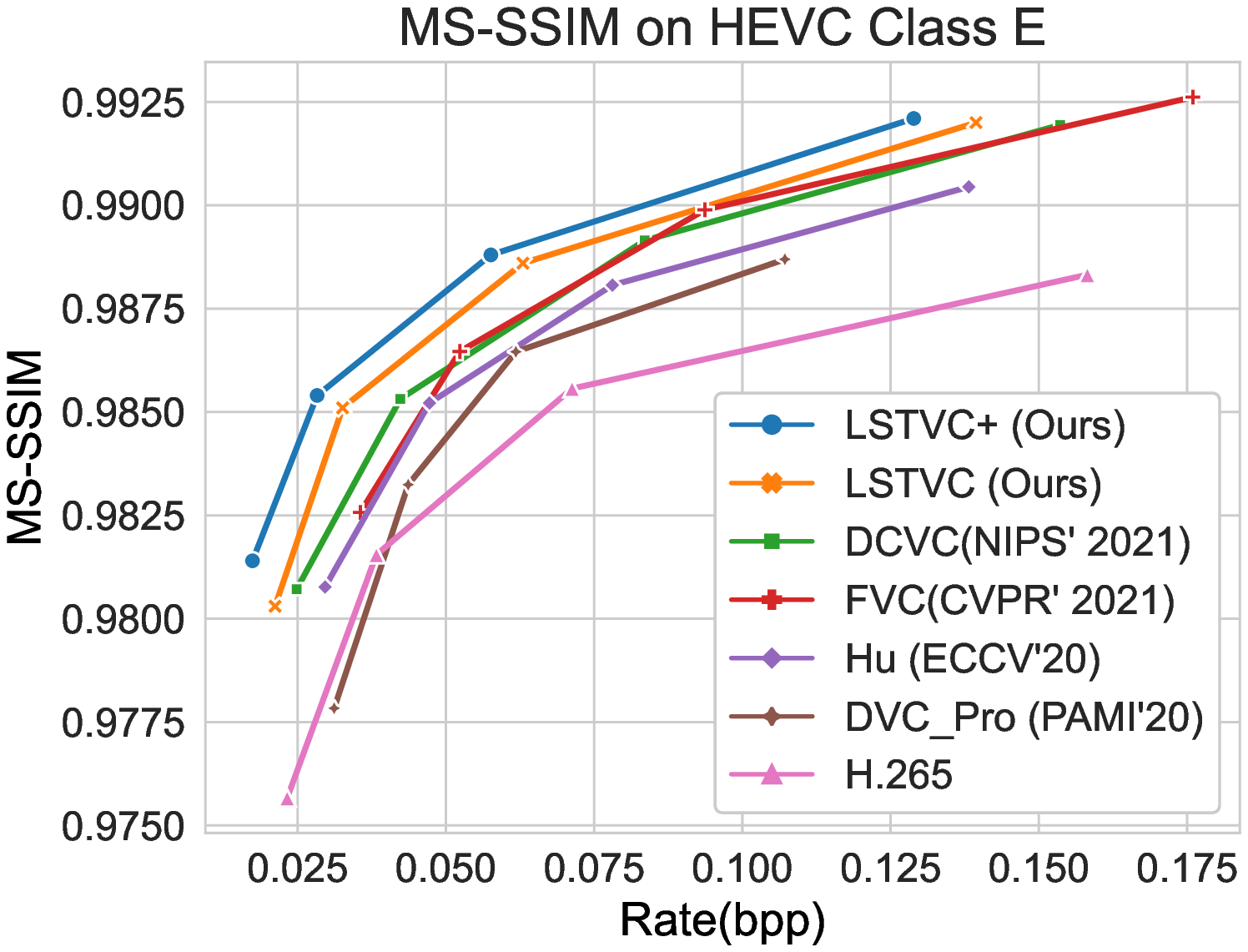}
			\label{ClassE_MSSSIM}
		\end{subfigure}
		\begin{subfigure}{.28\textwidth}
			\centering
			\includegraphics[width=\textwidth]{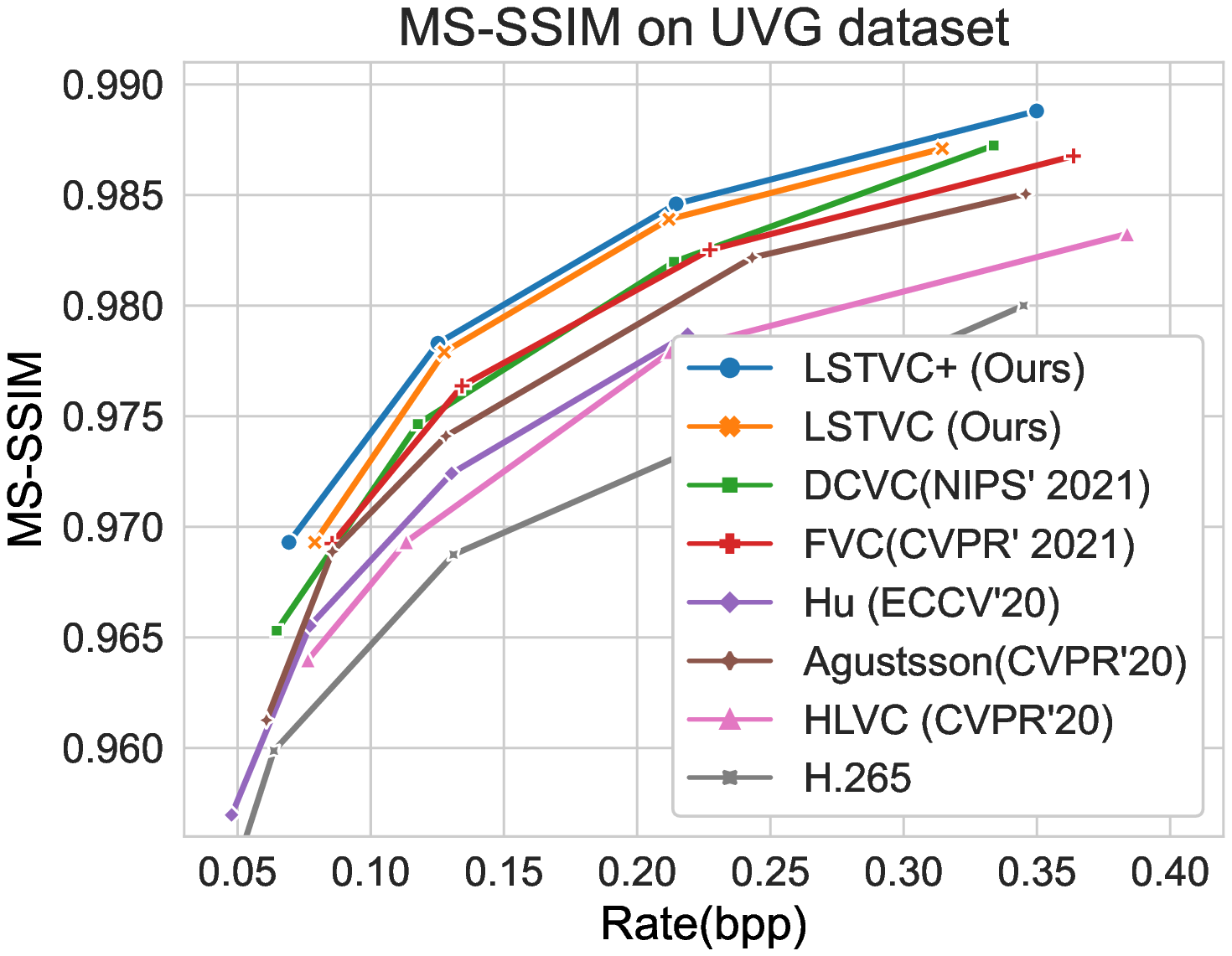}
			\label{UVG_MSSSIM}
		\end{subfigure}
		\begin{subfigure}{.28\textwidth}
			\centering
			\includegraphics[width=\textwidth]{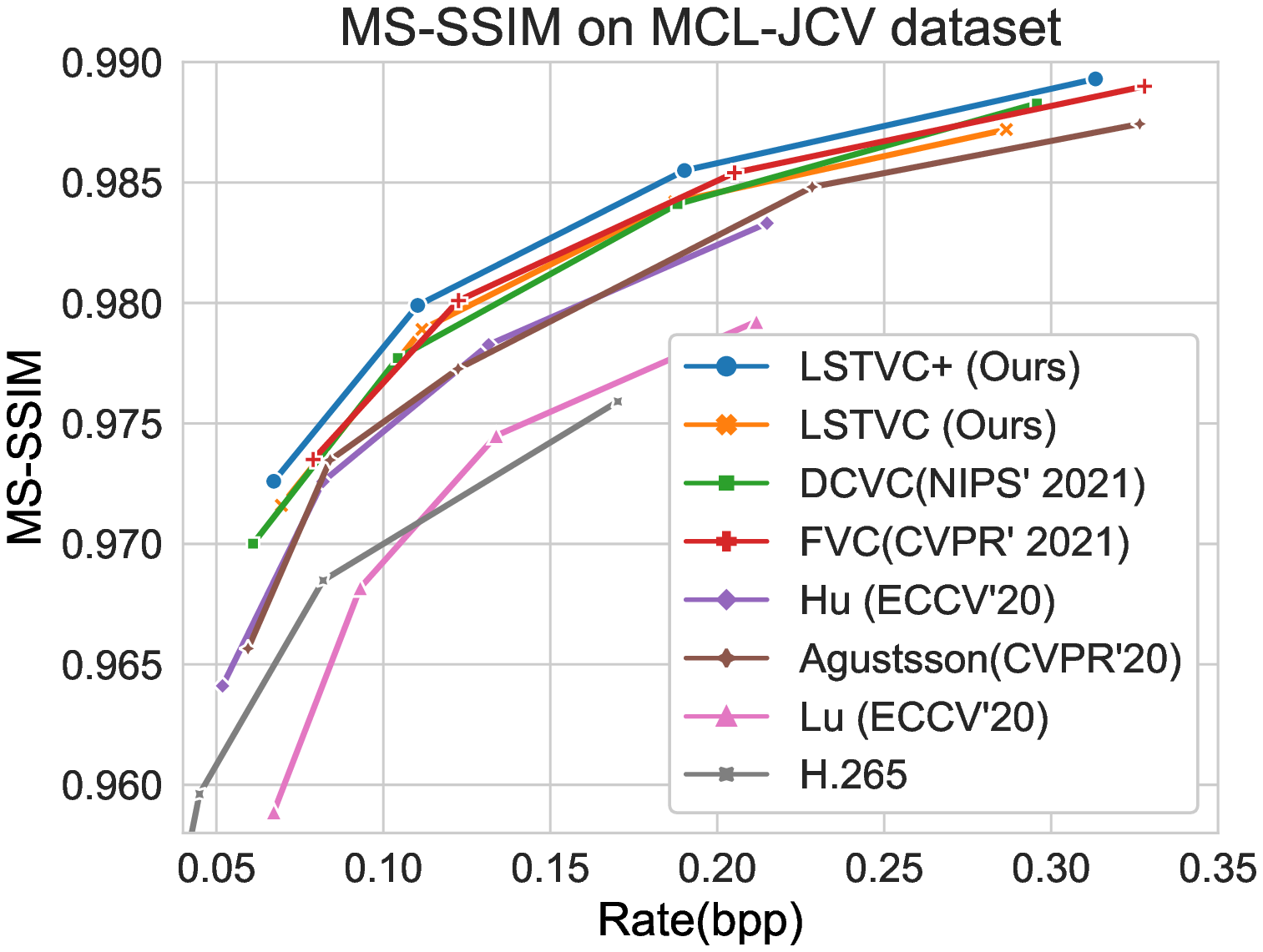}
			\label{MCLJCV_MSSSIM}
		\end{subfigure}
		\caption{The rate-distortion performance of our approach compared with H.265 (x265 LDP placebo) and the recent learned video compression approaches on the HEVC, UVG and MCL-JCV sequences.}
		\label{Result}
		\vspace{-10pt}
	\end{figure*}
	
	\section{Experiments}
	\subsection{Datasets and Implementation Details}
	\subsubsection{Datasets}
	Following previous work like \cite{lu2020end,li2021deep}, we adopt the Vimeo90K dataset provided by \cite{xue:2019}. The dataset contains 89,800 video clips with various and complex real-world motions, and each video sample contains seven consecutive frames. To validate the effectiveness of our model and compare performance with other state-of-the-art video compression approaches, we test our models on three datasets: the HEVC dataset \cite{bossen2013common}, the UVG dataset \cite{UVG} and the MCL-JCV dataset \cite{wang2016mcl}. The resolution of these datasets contains $416\times240$, $832\times480$, $1280\times720$ and $1920\times1080$.
	
	\subsubsection{Training settings}
	During training, our methods require the previous reconstructed frame $\hat{I}_{t-1} $ for motion compensation, so we train our model six times in succession using seven consecutive frames by buffering the previous reconstructed frame. The temporal prior is buffered in memory too. With regard to I frame compression, we choose cheng2020-anchor \cite{cheng2020learned} from CompressAI \cite{begaint2020compressai} to achieve higher rate-distortion performance than BPG \cite{bellard2014bpg}. 
		\begin{table*}[t]
		\caption{BDBR(\%) Results with the anchor of H.265 (x265 LDP placebo). \textbf{\textcolor{red}{Red}} and \textcolor{blue}{Blue} indicate the best and the second-best performance, respectively. The RD results of DVC, DVC\_Pro, HLVC and FVC are provided by their authors. The RD results of DCVC are calculated using the authors' official code.}
		\label{bdbr}
		\begin{center}
			\scalebox{0.81}{
				\begin{tabular}{c|cccccccccccccc}
					\toprule[1.2pt]
					&
					\multicolumn{7}{c|}{BDBR (\%) calculated by PSNR} &
					\multicolumn{7}{c}{BDBR (\%) calculated by MS-SSIM} \\ \cmidrule{2-15} 
					&
					\begin{tabular}[c]{@{}c@{}}DVC\\ \cite{Lu2019CVPR}\end{tabular} &
					\begin{tabular}[c]{@{}c@{}}DVC\_Pro\\ \cite{lu2020end}\end{tabular} &
					\begin{tabular}[c]{@{}c@{}}HLVC\\ \cite{yang2020learning}\end{tabular} &
					\begin{tabular}[c]{@{}c@{}}FVC\\ \cite{hu2021fvc}\end{tabular} &
					\begin{tabular}[c]{@{}c@{}}DCVC\\ \cite{li2021deep}\end{tabular} &
					\begin{tabular}[c]{@{}c@{}}LSTVC\\ (Ours)\end{tabular} &
					\multicolumn{1}{c|}{\begin{tabular}[c]{@{}c@{}}LSTVC+\\ (Ours)\end{tabular}} &
					\begin{tabular}[c]{@{}c@{}}DVC\\ \cite{Lu2019CVPR}\end{tabular} &
					\begin{tabular}[c]{@{}c@{}}DVC\_Pro\\ \cite{lu2020end}\end{tabular} &
					\begin{tabular}[c]{@{}c@{}}HLVC\\ \cite{yang2020learning}\end{tabular} &
					\begin{tabular}[c]{@{}c@{}}FVC\\ \cite{hu2021fvc}\end{tabular} &
					\begin{tabular}[c]{@{}c@{}}DCVC\\ \cite{li2021deep}\end{tabular} &
					\begin{tabular}[c]{@{}c@{}}LSTVC\\ (Ours)\end{tabular} &
					\begin{tabular}[c]{@{}c@{}}LSTVC+\\ (Ours)\end{tabular} \\ \cmidrule{2-15} 
					\multirow{-5}{*}{Dataset} &
					\multicolumn{9}{c|}{Optimized for PSNR} &
					\multicolumn{5}{c}{Optimized for MS-SSIM} \\ \midrule[1.2pt]
					\begin{tabular}[c]{@{}c@{}}HEVC\\  Class B\end{tabular} &
					25.55 &
					-0.24 &
					10.09 &
					-5.99 &
					-26.86 &
					\textcolor{blue}{-31.94} &
					\multicolumn{1}{c|}{\textcolor{red}{ \textbf{-35.89}}} &
					17.81 &
					-7.13 &
					-22.36 &
					-44.32 &
					-40.92 &
					\textcolor{blue}{ -50.31} &
					\textcolor{red}{ \textbf{-55.59}} \\ \midrule
					\begin{tabular}[c]{@{}c@{}}HEVC\\  Class C\end{tabular} &
					40.08 &
					13.16 &
					24.01 &
					-4.54 &
					-6.60 &
					\textcolor{blue}{-23.90} &
					\multicolumn{1}{c|}{\textcolor{red}{ \textbf{-30.39}}} &
					4.45 &
					-7.93 &
					-13.45 &
					-40.15 &
					-36.79 &
					\textcolor{blue}{ -50.61} &
					\textcolor{red}{ \textbf{-56.64}} \\ \midrule
					\begin{tabular}[c]{@{}c@{}}HEVC\\  Class D\end{tabular} &
					39.70 &
					20.86 &
					8.18 &
					-1.10 &
					-12.15 &
					\textcolor{blue}{ -26.23} &
					\multicolumn{1}{c|}{\textcolor{red}{ \textbf{-31.45}}} &
					4.56 &
					-8.44 &
					-24.05 &
					-43.54 &
					-42.14 &
					\textcolor{blue}{ -51.94} &
					\textcolor{red}{ \textbf{-57.53}} \\ \midrule
					\begin{tabular}[c]{@{}c@{}}HEVC\\ Class E\end{tabular} &
					14.72 &
					-6.54 &
					- &
					-6.21 &
					-28.93 &
					\textcolor{blue}{ -38.65} &
					\multicolumn{1}{c|}{\textcolor{red}{ \textbf{-43.07}}} &
					6.62 &
					-11.38 &
					- &
					-35.99 &
					-37.98 &
					\textcolor{blue}{ -48.78} &
					\textcolor{red}{ \textbf{-59.11}} \\ \midrule
					\begin{tabular}[c]{@{}c@{}}UVG\\ (1080p)\end{tabular} &
					26.73 &
					- &
					17.30 &
					-13.28 &
					-20.72 &
					\textcolor{blue}{ -26.19} &
					\multicolumn{1}{c|}{\textcolor{red}{ \textbf{-33.08}}} &
					32.30 &
					- &
					-20.55 &
					-44.39 &
					-43.05 &
					\textcolor{blue}{ -51.54} &
					\textcolor{red}{ \textbf{-55.08}} \\ \midrule
					\begin{tabular}[c]{@{}c@{}}MCL-JCV\\ (1080p)\end{tabular} &
					- &
					- &
					- &
					1.84 &
					-3.13 &
					\textcolor{blue}{ -17.72} &
					\multicolumn{1}{c|}{\textcolor{red}{ \textbf{-21.22}}} &
					- &
					- &
					- &
					\textcolor{blue}{-43.45} &
					-40.85 &
					-42.09 &
					\textcolor{red}{ \textbf{-48.05}} \\ \bottomrule[1.2pt]
				\end{tabular}
				\vspace{-10pt}
			}
		\end{center}
		
	\end{table*}
	
	We train eight models with different $\lambda $ to control RD performance ($\lambda $ = 256, 512, 1024, 2048 for the PSNR model and $\lambda $ = 8, 16, 32, 64 for the MS-SSIM model) for each framework. The batch size is set to 16, and we randomly crop the training sequences into the resolution of $256 \times 256$ as most video compression approaches do. The AdamW optimizer is adopted whose parameters $\beta_{1}$ and $\beta_{2}$ are set as $0.9$ and $0.999$. The learning rate is initialized to $1 \times 10^{-4}$ and decreased following the cosine annealing strategy. The entire network converges after 700000 iterations. All experiments are conducted using the PyTorch with NVIDIA RTX 3090 GPUs.

	\subsubsection{Test setting}
	When comparing with the current learned video compression methods, we set intra period to 10 for the HEVC dataset \cite{bossen2013common} and 12 for other datasets (the UVG \cite{UVG} and MCL-JCV \cite{wang2016mcl}) for fair comparisons. Besides, our framework requires the input frame's width and height to be a multiple of 64, so we crop the frame margin to meet the model requirements as previous work does.

	In order to compare rate-distortion performance with hybrid video coding standards \cite{HEVC}, we refer to the setting in \cite{Lu2019CVPR,yang2020learning} to generate the bitstream of x265 with FFmpeg \cite{x265}. Specifically, total frame number is set as 100 for the HEVC dataset. Following \cite{9288876}, QP is set to 15, 19, 23, 27 for the HEVC and MCL-JCV datasets, and set to 11, 15, 19, 23 for the UVG dataset.

	As mentioned above, to make comparison in more practical senarios, we also provide the RD results of x265, HM, VTM and DCVC when intra period is set to 32. In this test condition, we compress 96 frames for each sequence in the HEVC dataset following \cite{li2021deep}. For x265, it is simple to change intra period by changing \textit{keyint} parameter. As for HM-16.20, we use \textit{encoder\_lowdelay\_P\_main.cfg} from default configurations and set IntraPeriod to 32. For VTM-11.0, we use \textit{encoder\_lowdelay\_P\_vtm.cfg} and set IntraPeriod to 32 too. For these two reference software, the QP is set to 22,27,32,37. To get the results of DCVC \cite{li2021deep}, we use the official source codes of DCVC\footnote{\href{https://github.com/microsoft/DCVC}{https://github.com/microsoft/DCVC}} and change the GOP to 32.

	\subsection{Comparison with Other Methods}
	\subsubsection{Comparison methods and configuration}
	We compare our method with the recent video compression methods: DVC \cite{Lu2019CVPR}, DVC\_Pro \cite{lu2020end}, HLVC \cite{yang2020learning}, SSF \cite{agustsson2020scale}, Hu \cite{hu2020improving}, FVC \cite{hu2021fvc} and DCVC \cite{li2021deep}. As for the traditional method H.265 \cite{HEVC}, we refer to the command line in \cite{Lu2019CVPR,hu2020improving} and use x265 \cite{x265} with the LDP (Low Delay P) placebo mode. To be consistent with the test condition on \cite{Lu2019CVPR,yang2020learning,hu2020improving}, we test the HEVC dataset on the first 100 frames and test UVG, MCL-JCV sequences on all frames, and intra period is set to 10 for the HEVC dataset and 12 for other datasets. Moreover, to demonstrate the superiority of our method in long intra period conditions and compare to HM and VTM using a more reasonable configuration, we also provide the RD results of LSTVC and LSTVC+ when intra period is set to 32. Please refer to Table~\ref{BDBR_ip32} for more concrete results.
	
		\begin{table}[t]
		\caption{BDBR(\%) Results with the anchor of H.265 (x265 LDP placebo) \textbf{when intra period is set to 32}. \textbf{\textcolor{red}{Red}} and \textcolor{blue}{Blue} indicate the best and the second-best performance, respectively. The RD results of DCVC are calculated by the official open source codes.}
		\label{BDBR_ip32}
		\begin{center}
			\scalebox{0.88}{
				\begin{tabular}{cccccc}
					\toprule[1.2pt]  
					\multicolumn{6}{c}{BDBR (\%) calculated by PSNR} \\ \midrule
					\multicolumn{1}{l|}{Dataset} &
					\multicolumn{1}{l}{HM-16.20} &
					\multicolumn{1}{l}{VTM-11.0} &
					\multicolumn{1}{l}{\begin{tabular}[c]{@{}c@{}}DCVC\\ \cite{li2021deep}\end{tabular}} &
					\multicolumn{1}{l}{\begin{tabular}[c]{@{}c@{}}LSTVC\\ (Ours)\end{tabular}} &
					\multicolumn{1}{l}{\begin{tabular}[c]{@{}c@{}}LSTVC+\\ (Ours)\end{tabular}} \\ \midrule
					\multicolumn{1}{l|}{Class B} &
					-22.71 &
					\textcolor{red}{ \textbf{-45.28}} &
					-3.79 &
					-21.24 &
					\textcolor{blue}{ -28.60} \\
					\multicolumn{1}{l|}{Class C} &
					-19.47 &
					\textcolor{red}{ \textbf{-39.72}} &
					30.02 &
					-12.24 &
					\textcolor{blue}{ -24.45} \\
					\multicolumn{1}{l|}{Class D} &
					-18.50 &
					\textcolor{red}{ \textbf{-37.76}} &
					17.05 &
					-18.39 &
					\textcolor{blue}{ -27.28} \\
					\multicolumn{1}{l|}{Class E} &
					-26.14 &
					\textcolor{red}{ \textbf{-51.75}} &
					21.53 &
					-11.61 &
					\textcolor{blue}{ -28.12} \\ \midrule[1.2pt]
					\multicolumn{6}{c}{BDBR (\%) calculated by MS-SSIM} \\ \midrule
					\multicolumn{1}{l|}{Dataset} &
					\multicolumn{1}{l}{HM-16.20} &
					\multicolumn{1}{l}{VTM-11.0} &
					\multicolumn{1}{l}{\begin{tabular}[c]{@{}c@{}}DCVC\\ \cite{li2021deep}\end{tabular}} &
					\multicolumn{1}{l}{\begin{tabular}[c]{@{}c@{}}LSTVC\\ (Ours)\end{tabular}} &
					\multicolumn{1}{l}{\begin{tabular}[c]{@{}c@{}}LSTVC+\\ (Ours)\end{tabular}} \\ \midrule
					\multicolumn{1}{l|}{Class B} &
					-13.97 &
					\textcolor{blue}{ -49.29} &
					-31.08 &
					-44.71 &
					\textcolor{red}{ \textbf{-50.68}} \\
					\multicolumn{1}{l|}{Class C} &
					-14.49 &
					\textcolor{blue}{ -43.78} &
					-25.33 &
					-43.04 &
					\textcolor{red}{ \textbf{-50.14}} \\
					\multicolumn{1}{l|}{Class D} &
					-10.58 &
					-38.09 &
					-32.93 &
					\textcolor{blue}{ -43.68} &
					\textcolor{red}{ \textbf{-50.33}} \\
					\multicolumn{1}{l|}{Class E} &
					-18.62 &
					\textcolor{red}{ \textbf{-63.49}} &
					-18.34 &
					-29.85 &
					\textcolor{blue}{ -45.18} \\ \bottomrule[1.2pt]  
				\end{tabular}
			}
		\end{center}
		\vspace{-10pt}
	\end{table}
	
	\subsubsection{Rate-distortion performance}
	Figure-\ref{Result} shows the experimental results on the HEVC, UVG, and MCL-JCV sequences while taking PSNR and MS-SSIM as quality measurements. The Rate-Distortion performance of our method can surpass other learned methods and H.265 (x265 LDP placebo) by a large margin in terms of both PSNR and MS-SSIM. The conditional coding-based method DCVC\cite{li2021deep} uses auto-regressive prior and achieves quite high RD performance compared with previously learned video compression methods. From the BDBR results calculated by PSNR, DCVC can save an average of 18.63\% bits on the HEVC dataset compared to x265 LDP placebo, while the average bits saving for FVC\cite{hu2021fvc} is less than 5\%. As for our proposed methods, LSTVC and LSTVC+ can save 30.18\% and 35.20\% without introducing parallel-unfriendly operations. In other words, LSTVC and LSTVC+ can achieve superior RD performance while maintaining high efficiency. Taking all test datasets into account, the proposed LSTVC and LSTVC+ can save a further 11.04\% and 16.12\% of bits relative to DCVC. As for MS-SSIM oriented model, our methods can still achieve the best two performances around the learned video compression methods, as they do in terms of PSNR.
	
	To alleviate error propagation, most previously learned video compression methods (such as DVC, DVC\_Pro and FVC, etc.) set intra period to 10 or 12. From the perspective of practicality, this configuration is very unreasonable and unfair to the traditional methods. Therefore, we conduct extensive experiments on setting the intra period to 32 and take the standard reference software into account. Table~\ref{BDBR_ip32} shows the RD performance comparison when the intra period is set to 32. For the BDBR results calculated by PSNR, our methods can surpass the current SOTA compression method DCVC significantly with the merit of our temporal information utilization scheme. Besides, the performance of LSTVC+ can steadily exceed HM-16.20 on different tests, which previous video compression methods cannot achieve. From the results, our methods still has a gap with VTM-11.0 in PSNR, but in terms of MS-SSIM, LSTVC+ can surpass VTM-11.0 on most sequences.
	\begin{table}[t]
		\caption{BDBR(\%) Results on individual sequences with the anchor of H.265 (x265 LDP placebo) \textbf{when intra period is set to 32}. \textbf{\textcolor{red}{Red}} and \textcolor{blue}{Blue} indicate the best and the second-best performance, respectively.}
		\label{BDBR_sequence_ip32}
		\begin{center}
			\scalebox{0.85}{
				\begin{tabular}{ccccc}
					\toprule[1.2pt]
					\multicolumn{5}{c}{BDBR (\%) calculated by PSNR}                             \\ \midrule
					Dataset & Video & \begin{tabular}[c]{@{}c@{}}DCVC\\ \cite{li2021deep}\end{tabular} & \begin{tabular}[c]{@{}c@{}}LSTVC\\ (Ours)\end{tabular} & \begin{tabular}[c]{@{}c@{}}LSTVC+\\ (Ours)\end{tabular} \\ \midrule
					\multirow{5}{*}{\begin{tabular}[c]{@{}c@{}}HEVC\\ Class B\end{tabular}} & \textit{BasketballDrive } & -10.60 & \textcolor{blue}{-11.05} & \textcolor{red}{ \textbf{-17.30}} \\
					& \textit{BQTerrace }       & 25.63  & \textcolor{blue}{-10.49} & \textcolor{red}{ \textbf{-25.78}} \\
					& \textit{Cactus}           & 7.26   & \textcolor{blue}{-25.31} & \textcolor{red}{ \textbf{-35.17}} \\
					& \textit{Kimono }          & -29.03 & \textcolor{blue}{-33.82} & \textcolor{red}{ \textbf{-36.39}} \\
					& \textit{ParkScene }       & -0.50  & \textcolor{blue}{-24.73} & \textcolor{red}{ \textbf{-31.80}} \\ \midrule
					\multirow{4}{*}{\begin{tabular}[c]{@{}c@{}}HEVC\\ Class C\end{tabular}} & \textit{BasketballDrill}  & 22.77  & \textcolor{blue}{-15.81} & \textcolor{red}{ \textbf{-25.65}} \\
					& \textit{BQMall }          & 48.13  & \textcolor{blue}{-2.10}  & \textcolor{red}{ \textbf{-22.43}} \\
					& \textit{PartyScene }      & 43.94  & \textcolor{blue}{-7.10}  & \textcolor{red}{ \textbf{-17.93}} \\
					& \textit{RaceHorses} & 8.78   & \textcolor{blue}{-22.39} & \textcolor{red}{ \textbf{-32.63}} \\ \midrule
					\multirow{4}{*}{\begin{tabular}[c]{@{}c@{}}HEVC\\ Class D\end{tabular}} & \textit{BasketballPass}   & 19.75  & \textcolor{blue}{-9.57}  & \textcolor{red}{ \textbf{-19.45}} \\
					&\textit{ BlowingBubbles }  & -3.60  & \textcolor{blue}{-27.75} & \textcolor{red}{ \textbf{-35.91}} \\
					& \textit{BQSquare }        & 84.01  & \textcolor{blue}{1.89}   & \textcolor{red}{ \textbf{-13.21}} \\
					& \textit{RaceHorses }      & -12.70 & \textcolor{blue}{-32.45} & \textcolor{red}{ \textbf{-38.50}} \\ \midrule
					\multirow{3}{*}{\begin{tabular}[c]{@{}c@{}}HEVC\\ Class E\end{tabular}} & \textit{vidyo1}           & 30.86  & \textcolor{blue}{-9.21}  & \textcolor{red}{ \textbf{-25.95}} \\
					& \textit{vidyo3   }        & 19.16  & \textcolor{blue}{-16.22} & \textcolor{red}{ \textbf{-28.70}} \\
					& \textit{vidyo4 }          & 16.23  & \textcolor{blue}{-16.80} & \textcolor{red}{ \textbf{-29.56}} \\ \bottomrule[1.2pt]
				\end{tabular}
			}
		\end{center}
		\vspace{-10pt}
	\end{table}
	To understand performance variations according to individual test sequences and their characteristics, we provide the detailed RD performance of the individual HEVC sequences in Table~\ref{BDBR_sequence_ip32}. Compared to the SOTA method DCVC, our methods can achieve superior performance, especially on sequences with complex content and textures, such as \textit{BQTerrace}, \textit{PartyScene}, and \textit{BQSquare}. Besides, LSTVC and LSTVC+ perform much better in conference scenarios due to the advanced temporal information extraction capability.

	\subsection{Ablation Studies and Performance Analysis}
	\subsubsection{Temporal prior}To verify the effectiveness of the temporal prior, we conduct experiments by removing the temporal prior in motion compression and contextual compression, respectively. From Table~\ref{tab:ablation}, the temporal prior can bring in much more performance gain when using the baseline compensation method than when using the PGMC. It represents that the temporal prior also contains a wealth of short-range timing information due to its update strategy. Besides, adopting temporal prior can further save 12\% bits than the method only with PGMC, which means temporal prior contains valuable prior information that is missing from the short-range information of context prior. In other words, the two ideas in exploring long- and short-range temporal information are complementary. It makes sense that the gain from contextual compression is higher than that from motion compression because the temporal prior contains temporal information from previously decoded frames, so it has a higher spatial correlation with the current frame rather than the motion vectors.
	\begin{table*}[t]
		\centering
		\caption{The ablation study of the temporal prior. We provide the BDBR (\%) Results with the anchor of H.265 (x265 LDP placebo). \textbf{\textcolor{red}{Red}} indicate the best RD performance.}
		\label{tab:ablation}
		\begin{small}
			\scalebox{0.83}{\begin{tabular}{c|cc||cccc}
					\toprule[1.2pt]
					\multirow{2}{*}{} &
					\multicolumn{2}{c||}{Utilization of Long-range Temporal Information} &
					\multirow{3}{*}{\begin{tabular}[c]{@{}c@{}}HEVC\\ Class B\end{tabular}} &
					\multirow{3}{*}{\begin{tabular}[c]{@{}c@{}}HEVC\\ Class C\end{tabular}} &
					\multirow{3}{*}{\begin{tabular}[c]{@{}c@{}}HEVC\\ Class D\end{tabular}} &
					\multirow{3}{*}{\begin{tabular}[c]{@{}c@{}}HEVC\\ Class E\end{tabular}} \\\cmidrule{2-3}
					&
					\begin{tabular}[c]{@{}c@{}}Using Temporal Prior for \\ Contextual Compression\end{tabular} &
					\begin{tabular}[c]{@{}c@{}}Using Temporal Prior for \\ Motion Compression\end{tabular} &
					&
					&
					&
					\\\midrule[1.2pt]
					\multirow{3}{*}{\begin{tabular}[c]{@{}c@{}}Using Multiscale\\ Motion Compensation\\ (Baseline)\end{tabular}} &
					\XSolidBrush &\XSolidBrush &4.94 &19.63 &21.53 &15.35 \\
					&
					\Checkmark &\XSolidBrush &-12.21 &0.20 &	1.98 &-7.63 \\
					&
					\Checkmark &\Checkmark &-16.58 &		-4.45 &		-3.22 &	-9.09 \\\midrule
					\multirow{3}{*}{\begin{tabular}[c]{@{}c@{}}Using PGMC\\ (Ours)\end{tabular}} &
					\XSolidBrush &\XSolidBrush &-19.78 &-7.90 &		-11.78 &-16.21 \\
					&
					\Checkmark &	\XSolidBrush &-28.10 &-21.35 &-24.91 &-37.16 \\
					&
					\Checkmark &\Checkmark &\textcolor{red}{\textbf{-31.94}} &\textcolor{red}{\textbf{-23.90}} &\textcolor{red}{\textbf{-26.23}} &\textcolor{red}{\textbf{-38.65}} \\\bottomrule[1.2pt]
			\end{tabular}}
		\end{small}
		\vspace{-10pt}
	\end{table*}
	\begin{table}[t]
		\centering
		\caption{The detailed ablation study of the progressive guided motion compensation. We provide the BDBR (\%) Results with the anchor of H.265 (x265 LDP placebo). \textbf{\textcolor{red}{Red}} indicates the best RD performance.}
		\label{tab:PGMC_ablation}
		\setlength{\tabcolsep}{2mm}{
			\begin{small}
				\scalebox{0.80}{
					\begin{tabular}{c|cccc}
						\toprule[1.2pt]
						Compensation Operation                                                            & Class B  & Class C  & Class D  & Class E  \\ \midrule[1.2pt]
						\begin{tabular}[c]{@{}c@{}}Multi-scale flow warping\\ (w/o guidance)\end{tabular} & -9.24    & 0.40     & 0.34     & 1.21     \\ \midrule
						\begin{tabular}[c]{@{}c@{}}Guiding in \\ the smallest scale\end{tabular}          & -16.36 & -13.93 & -17.02 & -9.89  \\ \midrule
						\begin{tabular}[c]{@{}c@{}}Guiding in \\ the smallest two scales\end{tabular} & -21.65 & -17.35 & -21.73 & -29.20 \\ \midrule
						\begin{tabular}[c]{@{}c@{}}Guiding in all scales\\ (Ours)\end{tabular}            & \textcolor{red}{\textbf{-31.94}} & \textcolor{red}{\textbf{-23.90}} & \textcolor{red}{\textbf{-26.23}} & \textcolor{red}{\textbf{-38.65}} \\ \bottomrule[1.2pt]
				\end{tabular}}
			\end{small}
		}
		\vspace{-10pt}
	\end{table}

	\subsubsection{Progressive guided motion compensation (PGMC)}
	To evaluate the effectiveness of our proposed PGMC, we first replace the progressive guiding strategy with multi-scale optical flow warping under several conditions when different priors are provided. From Table~\ref{tab:ablation}, we can observe that using the PGMC will save 15\% to 20\% bits at the same PSNR when no temporal prior is used and save bits more than 20\% when the model is equipped with the temporal prior. 
	
	We also provide detailed ablation studies concerning the number of guiding layers. Since we design a multi-scale compensation structure, we demonstrate the effectiveness of the guiding scheme by adding guidance strategy scale by scale. From Table~\ref{tab:PGMC_ablation}, we find that the baseline using multi-scale flow warping (only $4\times$ and $1\times$ two-scale flow warping) without any guidance achieves about the same performance as x265 LDP placebo. By adding flow-to-kernel guidance on the smallest scale, the method can save about 14\% more bits compared to the baseline. When we add flow-to-kernel and scale-by-scale guidance to the smallest two scales, we can save 22\% bits with the anchor of x265 LDP placebo. Regarding the complete model, our PGMC operation can save more than 30\% bits compared to the baseline. These results demonstrate the advantage of our proposed modules.

	\subsubsection{Rate-distortion performance}
	Fig.~\ref{Result} shows the experimental results on the HEVC, UVG, and MCL-JCV sequences while taking PSNR and MS-SSIM as quality measurements. The Rate-Distortion performance of our method can surpass other learned methods and H.265 (x265 LDP placebo) by a large margin in terms of both PSNR and MS-SSIM. The conditional coding-based method DCVC \cite{li2021deep} uses auto-regressive prior and achieves quite high RD performance compared with previously learned video compression methods. From the BDBR results calculated by PSNR, DCVC can save an average of 18.63\% bits on the HEVC dataset compared to x265 LDP placebo, while the average bits saving for FVC \cite{hu2021fvc} is less than 5\%. As for our proposed methods, LSTVC and LSTVC+ can save 30.18\% and 35.20\% without introducing parallel-unfriendly operations. In other words, LSTVC and LSTVC+ can achieve superior RD performance while maintaining high efficiency. Taking all test datasets into account, the proposed LSTVC and LSTVC+ can save a further 11.04\% and 16.12\% of bits relative to DCVC. As for MS-SSIM oriented model, our methods can still achieve the best two performances around the learned video compression methods, as they do in terms of PSNR.
		
	\subsubsection{Model complexity and running time}
	The parameters of the LSTVC and LSTVC+ are about 32 and 49 megabytes, respectively.  Table~\ref{tab:speed} shows the detailed inference time of DCVC and our proposed methods. We take the sequences in the HEVC dataset to test inference time. We record the inference time of coding, including network forward and the entropy model. Note that DCVC utilizes auto-regression to capture spatial redundancy, causing its inference time to take a long time by the element-by-element coding order. Even though auto-regressive prior can further boost performance, we do not choose it since it effect inference speed dramatically. It can be seen that without parallel-unfriendly operation like auto-regressive entropy model, our proposed methods can achieve high efficient compression while maintaining promising RD performance. We also use MACs (multiply–accumulate operations) to evaluate the model complexity. We can conclude that our methods achieve fast inference speed through the introduction of more powerful transform networks, progressive guided motion compensation and temporal prior, all within a model designed to facilitate parallel computing.
	\begin{table}[t]
		\centering
		\caption{The inference time cost per frame and complexity of the DCVC, LSTVC and LSTVC+ on the HEVC dataset. The time of encoding and decoding includes network inference time and entropy model running time. The MACs are calculated using 1080p frame.}
		\label{tab:speed}
		\setlength{\tabcolsep}{2mm}{
			\begin{small}
				\scalebox{0.79}{
					\begin{tabular}{cc|cccc|c}
						\toprule[1.2pt]
						\multicolumn{2}{c|}{} &
						\begin{tabular}[c]{@{}c@{}}HEVC\\ Class B\end{tabular} &
						\begin{tabular}[c]{@{}c@{}}HEVC\\ Class C\end{tabular} &
						\begin{tabular}[c]{@{}c@{}}HEVC\\ Class D\end{tabular} &
						\begin{tabular}[c]{@{}c@{}}HEVC\\ Class E\end{tabular} &
						MACs \\ \midrule[1.2pt]
						\multicolumn{1}{c|}{\multirow{2}{*}{DCVC \cite{li2021deep}}}  & Encoding & 16.41s & 3.31s & 0.66s & 6.95s & \multirow{2}{*}{2262 G} \\
						\multicolumn{1}{c|}{}                        & Decoding & 67.68s & 13.36s & 2.79s & 28.52s &                               \\ \midrule
						\multicolumn{1}{c|}{\multirow{2}{*}{LSTVC}}  & Encoding & 0.80s & 0.28s & 0.19s & 0.44s & \multirow{2}{*}{2969 G} \\
						\multicolumn{1}{c|}{}                        & Decoding & 0.47s & 0.15s & 0.10s & 0.26s &                               \\ \midrule
						\multicolumn{1}{c|}{\multirow{2}{*}{LSTVC+}} & Encoding & 0.89s & 0.30s & 0.20s & 0.49s & \multirow{2}{*}{4547 G} \\
						\multicolumn{1}{c|}{}                        & Decoding & 0.57s & 0.17s & 0.10s & 0.30s &                               \\ \bottomrule[1.2pt]
				\end{tabular}}
			\end{small}
		}
		\vspace{-10pt}
	\end{table}
	\begin{figure*}[t]
		\centering
		\begin{subfigure}{.29\textwidth}
			\centering
			\includegraphics[width=\linewidth]{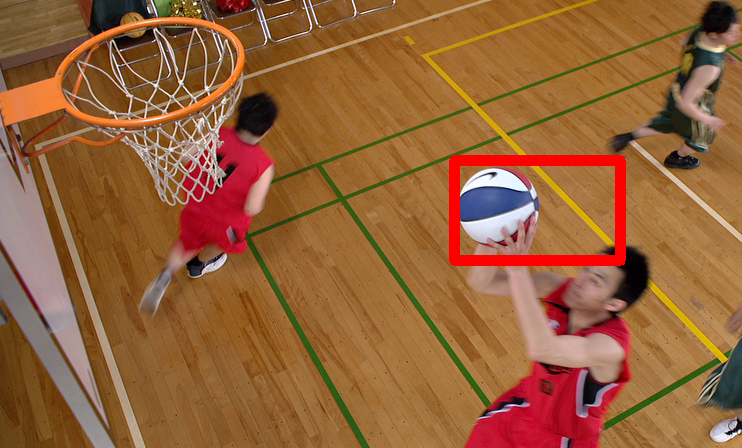}
			\label{BasketballDrive}
			\vspace{-10pt}
			\caption{HEVC Class C \protect \\ BasketballDrill}
		\end{subfigure}
		\begin{subfigure}{.29\textwidth}
			\centering
			\includegraphics[width=\linewidth]{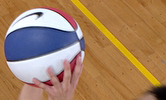}
			\label{BasketballDrive_GT}
			\vspace{-10pt}
			\caption{Ground Truth \protect \\(PSNR(dB)/BPP)}
		\end{subfigure}
		\begin{subfigure}{.29\textwidth}
			\centering
			\includegraphics[width=\linewidth]{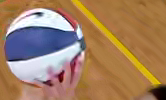}
			\label{BasketballDrive_x265_LDP_placebo}
			\vspace{-10pt}
			\caption{x265 LDP placebo  \protect \\(28.29/0.0692)}
		\end{subfigure}
		\\
		\begin{subfigure}{.29\textwidth}
			\centering
			\includegraphics[width=\linewidth]{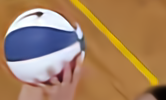}
			\label{BasketballDrive_DCVC}
			\vspace{-10pt}
			\caption{DCVC (PSNR)\protect \\(30.244/0.0537)}
		\end{subfigure}
		\begin{subfigure}{.29\textwidth}
			\centering
			\includegraphics[width=\linewidth]{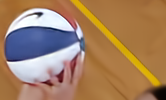}
			\label{BasketballDrive_LSTVC}
			\vspace{-10pt}
			\caption{LSTVC (PSNR) \protect \\(30.400/0.0482)}
		\end{subfigure}
		\begin{subfigure}{.29\textwidth}
			\centering
			\includegraphics[width=\linewidth]{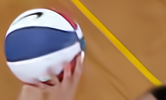}
			\label{BasketballDrive_LSTVC+}
			\vspace{-10pt}
			\caption{LSTVC+ (PSNR) \protect \\(30.457/0.0442)}
		\end{subfigure}
		\begin{subfigure}{.29\textwidth}
			\centering
			\includegraphics[width=\linewidth]{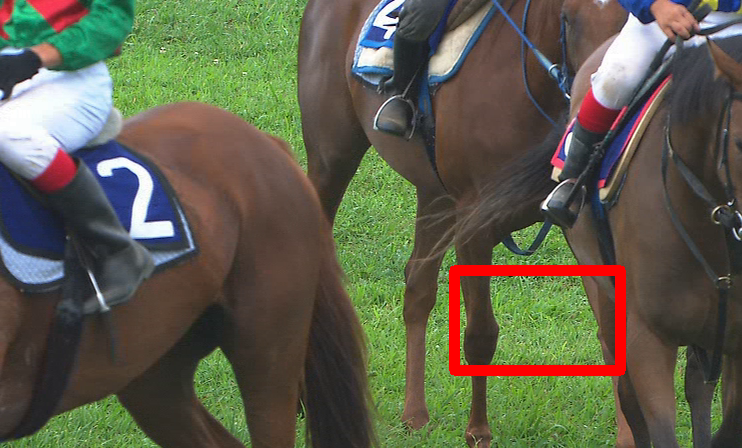}
			\caption{HEVC Class C \protect \\ RaceHorses}
		\end{subfigure}
		\begin{subfigure}{.29\textwidth}
			\centering
			\includegraphics[width=\linewidth]{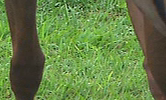}
			\caption{Ground Truth\protect \\(MS-SSIM/BPP)}
		\end{subfigure}
		\begin{subfigure}{.29\textwidth}
			\centering
			\includegraphics[width=\linewidth]{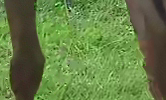}
			\caption{x265 LDP placebo\protect \\(0.9318/0.1141)}
		\end{subfigure}
		\\
		\begin{subfigure}{.29\textwidth}
			\centering
			\includegraphics[width=\linewidth]{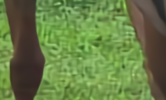}
			\caption{DCVC (MS-SSIM)\protect \\(0.9530/0.1087)}
		\end{subfigure}
		\begin{subfigure}{.29\textwidth}
			\centering
			\includegraphics[width=\linewidth]{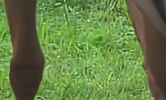}
			\caption{LSTVC (MS-SSIM)\protect \\(0.9618/0.1124)}
		\end{subfigure}
		\begin{subfigure}{.29\textwidth}
			\centering
			\includegraphics[width=\linewidth]{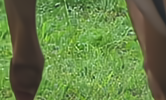}
			\caption{LSTVC+ (MS-SSIM)\protect \\(0.9639/0.1058)}
		\end{subfigure}
		\caption{The visual results of traditional hybrid codecs x265, DCVC \cite{li2021deep} and our LSTVC/LSTVC+. PSNR and MS-SSIM in brackets denote that the models are optimized for PNSR and MS-SSIM, respectively.}
		\label{VisualResult}
	\end{figure*}
	\subsection{Visualization results}
	This section presents the visual quality comparison of H.265 and our proposed models. We select four representative sequences in the HEVC dataset, including \textit{BasketballDrive} and \textit{RaceHorses}. Fig.~\ref{VisualResult} shows the visual quality comparison of the H.265 (x265 LDP placebo), DCVC, LSTVC and LSTVC+. With fewer bits consumed, the proposed methods produce fewer compression artifacts, less color shift and achieve higher subjective quality than the traditional hybrid codecs H.265. As for the comparison with DCVC, our proposed methods can reconstruct clearer and more correct textures. For instance, from (d), (e) and (f) in Fig.~\ref{VisualResult} we see that our methods can reconstruct a clearer outline of the basketball and the corresponding correct color fill. From (j), (k) and (l) in Fig.~\ref{VisualResult}, we find that LSTVC and LSTVC+ can recover more high-fidelity lawn texture than other methods.
	
	\section{Conclusion}
	This paper presents a learned video compression framework that focuses on exploring various temporal information. The proposed methods LSTVC and LSTVC+ can mine the unique characteristic of video content and effectively utilize long- and short-range temporal information. Specifically, for long-range temporal information exploitation, we propose the temporal prior that can be updated by decoded frame and motion during encoding and decoding. In that case, the temporal prior contains temporal information from all decoded video frames and can assist in parameter prediction and frame transformation. As for exploring short-range temporal information, we propose progressive guided motion compensation to achieve robust and effective compensation. In detail, we design a hierarchical structure to achieve multi-scale compensation. On the basis of this, we generate offsets at each scale using optical flow guidance, and smaller-scale compensation results will be used to guide the next scale's compensation. The two proposed modules are complementary to each other and help to improve the framework significantly. Experimental results demonstrate that our method can obtain better RD performance than the state-of-the-art video compression approaches. 
	
	
	\bibliographystyle{IEEEtran}
	\bibliography{egbib}
	\leavevmode
	\newline
	\textbf{Huairui Wang} is pursuing the Ph.D. degree at Wuhan University, China.
	\\
	\\
	\textbf{Zhenzhong Chen} is a Professor at Wuhan University, China.
	
\end{document}